\documentstyle[amstex,12pt]{article}
\begin{document}

\author{{\bf Fabio Cardone}$^{a,b}${\bf , Alessio Marrani}$^{c}${\bf \ and } \and
{\bf Roberto Mignani}$^{c,d}$ \\
$a$ Istituto per lo Studio dei Materiali Nanostrutturati\\
(ISMN-CNR)\\
Via dei Taurini, 19\\
00185 ROMA, Italy\\
$b$ I.N.D.A.M. - G.N.F.M.\\
$c$ Dipartimento di Fisica ''E. Amaldi''\\
Universit\`{a} degli Studi ''Roma Tre''\\
Via della Vasca Navale, 84\\
00146 ROMA, Italy\\
$d$ I.N.F.N. - Sezione di Roma III}
\title{{\bf Killing symmetries of generalized Minkowski spaces.}\\
{\bf 2- Finite structure of space-time rotation groups in four dimensions}}
\maketitle

\begin{abstract}
In this paper, we continue the study of the Killing symmetries of a
N-dimensional generalized Minkowski space, i.e. a space endowed with a (in
general non-diagonal) metric tensor, whose coefficients do depend on a set
of non-metrical coordinates. We discuss here the finite structure of the
space-time\ rotations in such spaces, by confining ourselves (without loss
of generality) to the four-dimensional case. In particular, the results
obtained are specialized to the case of a ''deformed'' Minkowski space $%
\widetilde{M_{4}}$ (i.e. a pseudoeuclidean space with metric coefficients
depending on energy), for which we derive the explicit general form of the
finite rotations and boosts in different parametric bases.
\end{abstract}

\bigskip

\section{INTRODUCTION\protect\bigskip}

In a previous paper [1], we started the study of the Killing symmetries of a
$N$-dimensional generalized\bigskip\ Minkowski space, i.e. a space endowed
with a (in general non-diagonal)\ \bigskip metric tensor, whose coefficients
do depend on a set of non-metrical coordinates. \bigskip In
particular,\bigskip\ we discussed the infinitesimal-algebraic structure of
the space-time\bigskip\ rotations in such a space.\

An example of a generalized Minkowski space is provided by the deformed
space-time $\widetilde{M_{4}}$ of Deformed Special Relativity (DSR). DSR is
a generalization of the standard Special Relativity (SR) based on a\
''deformation'' of space-time, assumed\bigskip\ to be endowed with a metric
whose coefficients depend on the energy of\bigskip\ the process considered
[2]. Such a formalism applies in principle to {\em all} four interactions
(electromagnetic, weak, strong and\bigskip\ gravitational) --- at least as
far as their\bigskip\ nonlocal behavior and\ nonpotential part is concerned
--- and provides a metric representation of them (at\ least for\bigskip\ the
process and in the\ energy range considered) ([3]-[6]). Moreover, it was
shown\bigskip\ that DSR is actually a\ five-dimensional scheme, in the\
sense that\bigskip\ the deformed Minkowski space can be naturally embedded
in\medskip\ a larger Riemannian manifold,\bigskip\ with energy as fifth
dimension [7].\bigskip\

In this paper, \ following the line of mathematical-formal research strated
by [1], [10] and [13], we shall continue our investigation by discussing the
finite structure of the space-time\bigskip\ rotations in generalized
Minkowski spaces. For simplicity's sake, we shall restrict us (without loss
of generality) to the four-dimensional case, by specializing our results to
the deformed space-time $\widetilde{M_{4}}$ of DSR.

The organization of the paper is as follows. In Sect. 2 we briefly review
the\bigskip\ formalism of DSR \ and of the deformed Minkowski space $%
\widetilde{M_{4}}$. The results obtained in [1], concerning the maximal
Killing group of generalized\bigskip\ $N$-dimensional\ Minkowski spaces, are
summarized in Subsect. 3.1, where we also give the general form of the
finite spacetime rotations in terms of the infinitesimal generators of the $%
N $-d. generalized, homogeneous Lorentz group. In Subsect. 3.2 the latter
results are specialized to the case of the\ deformed, homogeneous\ Lorentz
group $SO(3,1)_{DEF.}$ of DSR. Sect. 4 deals with the finite deformed
spacetime rotations about a coordinate axis. The explicit form of the
infinitesimal generators in the DSR case is given in Subsect. 4.1.
Subsections 4.2 and 4.3 discuss finite deformed 3-d. rotations about a
coordinate axis and finite deformed boosts along a coordinate axis,
respectively. Sect. 5 generalizes the latter results to the case of a
generic direction. Finally, Sect. 6 concludes the paper.\pagebreak

\section{\protect\bigskip DEFORMED\ SPECIAL RELATIVITY IN FOUR DIMENSIONS
(DSR4)}

The generalized (``deformed'') Minkowski space $\widetilde{M_{4}}$ (DMS4) of
DSR4\footnote{%
In the following, we shall use the more explicit notation DSR4 instead of
DSR, in order to stress the dimensionality of the spacetime involved, and to
distinguish it from the {\em 5-d. Deformed Relativity} (DR5), i.e. the
embedding of DSR in a 5-dimensional, truly Riemannian, manifold [7].} is
defined as a\bigskip\ space with the same local coordinates $x$ of $M_{4}$
(the four-vectors of the usual\bigskip\ Minkowski space), but with metric
given by the metric tensor\footnote{%
In the following, Greek indices vary in the range $\left\{ 0,1,2,3\right\} $%
, 3-vectors are denoted in bold and the notation ''ESC on'' (''ESC off'')
means that the Einstein sum convention on repeated indices is (is not)
understood.}\bigskip\
\begin{gather}
g_{\mu \nu
,DSR4}(x^{5})=diag(b_{0}^{2}(x^{5}),-b_{1}^{2}(x^{5}),-b_{2}^{2}(x^{5}),-b_{3}^{2}(x^{5}))=
\nonumber \\
\nonumber \\
\stackrel{\text{{\footnotesize ESC off}}}{=}\delta _{\mu \nu }\left[
b_{0}^{2}(x^{5})\delta _{\mu 0}-b_{1}^{2}(x^{5})\delta _{\mu
1}-b_{2}^{2}(x^{5})\delta _{\mu 2}-b_{3}^{2}(x^{5})\delta _{\mu 3}\right] ,
\end{gather}
\bigskip where the $\left\{ b_{\mu }^{2}(x^{5})\right\} $ are dimensionless,
real, positive functions of the \bigskip independent, non-metrical (n.m.)
variable $x^{5}$ \footnote{%
Such a coordinate is to be interpreted as the energy (see Refs. [3]-[7]);
moreover, the index $5$ explicitly refers to the above-mentioned fact that
the deformed Minkowski space can be {\em ''naturally'' embedded} in a
5-dimensional (Riemannian) space [7].}. The generalized interval in $%
\widetilde{M_{4}}$ is\bigskip\ therefore given by ($x^{\mu
}=(x^{0},x^{1},x^{2},x^{3})=(ct,x,y,z)$, with $c$ being the usual\bigskip\
light speed in vacuum)\bigskip\
\begin{gather}
ds^{2}(x^{5})=b_{0}^{2}(x^{5})c^{2}dt^{2}-(b_{1}^{2}(x^{5})dx^{2}+b_{2}^{2}(x^{5})dy^{2}+b_{3}^{2}(x^{5})dz^{2})=
\nonumber \\
\nonumber \\
=g_{\mu \nu ,DSR4}(x^{5})dx^{\mu }dx^{\nu }=dx\ast dx.
\end{gather}
\bigskip

The last step in (2) defines the ($x^{5}$-dependent) scalar product $\ast $
in the deformed Minkowski\bigskip\ space $\widetilde{M_{4}}$ .\bigskip\ In
order to emphasize the dependence of DMS4 on the variable $x^{5}$, we shall
sometimes use the notation $\widetilde{M_{4}}(x^{5})$. It follows
immediately \bigskip that $\widetilde{M_{4}}(x^{5})$ can be regarded as a
particular case of a Riemann space with null curvature.\bigskip

From the general condition
\begin{equation}
g_{\mu \nu ,DSR4}(x^{5})g_{DSR4}^{\nu \rho }(x^{5})=\delta _{\mu }^{~~\rho }
\end{equation}
we get for the contravariant components of the metric tensor
\begin{gather}
g_{DSR4}^{\mu \nu
}(x^{5})=diag(b_{0}^{-2}(x^{5}),-b_{1}^{-2}(x^{5}),-b_{2}^{-2}(x^{5}),-b_{3}^{-2}(x^{5}))=
\nonumber \\
\nonumber \\
\stackrel{\text{{\footnotesize ESC off}}}{=}\delta ^{\mu \nu }\left(
b_{0}^{-2}(x^{5})\delta ^{\mu 0}-b_{1}^{-2}(x^{5})\delta ^{\mu
1}-b_{2}^{-2}(x^{5})\delta ^{\mu 2}-b_{3}^{-2}(x^{5})\delta ^{\mu 3}\right) .
\end{gather}
\bigskip

Let us stress that metric (1) is supposed to hold at a {\em local} (and not
global)\bigskip\ scale. We shall therefore refer to it as a ``{\em topical}%
'' deformed metric, because it\bigskip\ is supposed to be valid not
everywhere, but only in a suitable (local) \bigskip space-time region
(characteristic of both the system and the interaction considered).\bigskip\
\ In brief, DSR4 can be regarded as a (local) {\em ''anisotropizing
deforming''} generalization of SR, synthetically expressed by the metric
transition $g_{\mu \nu ,SR}\rightarrow g_{\mu \nu ,DSR4}(x^{5})$ (with $%
g_{\mu \nu ,SR}\equiv diag\left( 1,-1,-1,-1\right) $).

The two basic postulates of DSR4 (which generalize those of
standard\bigskip\ SR) are [2]:\bigskip

\ \ 1- {\em Space-time properties: } Space-time is homogeneous, but space is
not\bigskip\ necessarily isotropic; a reference frame in which space-time is
endowed with\bigskip\ such properties is called a ''{\em topical'' reference
frame} (TIRF). Two TIRF's are\bigskip\ in general moving uniformly with
respect to each other (i.e., as in SR, they\bigskip\ are connected by a
''inertiality'' relation, which defines an equivalence class\bigskip\ of $%
\infty ^{3}$ TIRF );\bigskip

2- {\em Generalized Principle of Relativity }(or {\em Principle of Metric
Invariance}):\bigskip\ All physical measurements within each TIRF must be
carried out via the\bigskip\ {\em same} metric.\bigskip

The metric (1) is just a possible realization of the above postulates.
We\bigskip\ refer the reader to Refs. [3]-[7] for the explicit expressions
of the \bigskip phenomenological energy-dependent metrics for the four
fundamental interactions\footnote{%
Since the metric coefficients $b_{\mu }^{2}(x^{5})$ are {\em dimensionless},
they actually do depend on the ratio $\frac{x^{5}}{x_{0}^{5}}$, where
\[
x_{0}^{5}\equiv \ell _{0}E_{0}
\]
is a {\em fundamental length}, proportional (by the {\em %
dimensionally-transposing} constant $\ell _{0}$) to the {\em threshold energy%
} $E_{0}$, characteristic of the interaction considered (see Refs. [3]-[7]).}%
.\bigskip \pagebreak\

\section{GENERAL\ FORM\ OF\ FINITE\ SPACE-TIME\ ROTATIONS\ IN\
4-DIMENSIONAL\ GENERALIZED\ MINKOWSKI\ SPACES}

\subsection{Maximal\ Killing\ Group\ of\ Generalized Minkowski Spaces}

A $N$-dimensional {\em generalized Minkowski space }$\widetilde{M_{N}}%
(\left\{ x\right\} _{n.m.})$ is a Riemann space endowed with the global
metric structure [1]
\begin{equation}
ds^{2}=g_{\mu \nu }(\left\{ x\right\} _{n.m.})dx^{\mu }dx^{\nu },
\end{equation}
where the (in general non-diagonal) metric tensor $g_{\mu \nu }(\left\{
x\right\} _{n.m.})$ ($\mu ,\nu =1,2,...,N$) depends on a set $\left\{
x\right\} _{n.m.}$ of $N_{n.m.}$ non-metrical coordinates (i.e. different
from the $N$ coordinates related to the dimensions of the space considered).
We shall assume the (not necessarily hyperbolic) signature $(T,S)$ (i.e. $T$
timelike dimensions and $S=N-T$ spacelike dimensions). It follows that $%
\widetilde{M_{N}}(\left\{ x\right\} _{n.m.})$ is {\em flat}, because all the
components of the Riemann-Christoffel tensor vanish.

The maximal Killing group of $\widetilde{M_{N}}(\left\{ x\right\} _{n.m.})$
is the {\em generalized Poincar\'{e}} (or {\em inhomogeneous Lorentz}) {\em %
group} $P(S,T)_{GEN.}^{N(N+1)/2}$

\begin{equation}
P(T,S)_{GEN.}^{N(N+1)/2\text{{\footnotesize \ }}}=SO(T,S)_{GEN.}^{N(N-1)/2}%
\otimes _{s}Tr.(T,S)_{GEN.}^{N\text{{\small \ }}},
\end{equation}
i.e. the (semidirect) product of the Lie group of $N$-dimensional space-time
rotations (or $N$-d. generalized, homogeneous Lorentz group $%
SO(T,S)_{GEN.}^{N(N-1)/2\text{{\footnotesize \ }}}$) with $N(N-1)/2$
parameters, and of the Lie group of $N$-dimensional space-time translations $%
Tr.(T,S)_{GEN.}^{N}$ with $N$ parameters (see Ref. [1]). The semidirect [8]
nature of the group product is due to the fact that, as it shall be
explicitely derived (in the hyperbolically-signed case $N=4,S=3,T=1$\ of
DSR4, without loss of generality) in a forthcoming paper [9], in general we
have that
\begin{gather}
\exists \text{ at least 1 }\left( \mu ,\nu ,\rho \right) \in \left\{
1,2,...,N\right\} ^{3}:  \nonumber \\
\nonumber \\
:[I_{GEN.}^{\mu \nu }(\left\{ x\right\} _{n.m.}),\Upsilon _{GEN.}^{\rho
}(\left\{ x\right\} _{n.m.})]\neq 0,\text{ }\forall \left\{ x\right\}
_{n.m.},
\end{gather}
where $I_{GEN.}^{\mu \nu }(\left\{ x\right\} _{n.m.})$\ and $\Upsilon
_{GEN.}^{\rho }(\left\{ x\right\} _{n.m.})$\ are the infinitesimal
generators of $SO(T,S)_{GEN.}^{N(N-1)/2}$\ and $Tr.(T,S)_{GEN.}^{N\text{%
{\small \ }}}$, respectively.

The infinitesimal generators $\left\{ (I^{\alpha \beta })_{~~\nu }^{\mu
}(\left\{ x\right\} _{n.m.})\right\} _{\alpha ,\beta =1...N}$ of $\
SO(T,S)_{GEN.}^{N(N-1)/2}$ satisfy the {\em generalized Lorentz algebra} [1]
\begin{gather}
\lbrack I^{\alpha \beta }(\left\{ x\right\} _{n.m.}),I^{\rho \sigma
}(\left\{ x\right\} _{n.m.})]=  \nonumber \\
\nonumber \\
=g^{\alpha \sigma }(\left\{ x\right\} _{n.m.})I^{\beta \rho }(\left\{
x\right\} _{n.m.})+g^{\beta \rho }(\left\{ x\right\} _{n.m.})I^{\alpha
\sigma }(\left\{ x\right\} _{n.m.})+  \nonumber \\
\nonumber \\
-g^{\alpha \rho }(\left\{ x\right\} _{n.m.})I^{\beta \sigma }(\left\{
x\right\} _{n.m.})-g^{\beta \sigma }(\left\{ x\right\} _{n.m.})I^{\alpha
\rho }(\left\{ x\right\} _{n.m.}).
\end{gather}

On the basis of the results of [1], the infinitesimal transformation
corresponding, in a 4-d. generalized Minkowski space (with $S(<4)$ spacelike
and $T=4-S$ timelike dimensions) to the element $g$ of the generalized,
homogeneous Lorentz group $SO(S,T=4-S)_{GEN.}$, is given by\bigskip
\footnote{%
For precision's sake, in general for infinitesimal deformed boost
transformations $g\in SO(3,1)_{DEF.}$ should be replaced by $\delta g$,
where $\delta g\in su(2)_{DEF.}\times su(2)_{DEF.}$, i.e. it is an element
of the deformed Lorentz algebra. But, for simplicity's sake, we will omit,
but mean, this cumbersome notation.}\
\begin{gather}
\delta g:x^{\mu }\rightarrow \left( x^{\prime }\right) _{(g)}^{\mu }(\left\{
x\right\} _{m.},\left\{ x\right\} _{n.m.})=\left( x^{\mu }\right)
_{(g)}^{^{\prime }}(\left\{ x\right\} _{m.},\left\{ x\right\} _{n.m.})=
\nonumber \\
\nonumber \\
=x^{\mu }+\delta x_{(g)}^{\mu }(\left\{ x\right\} _{m.},\left\{ x\right\}
_{n.m.})=x^{\mu }+\delta \omega _{~~~\nu }^{\mu }(g,\left\{ x\right\}
_{n.m.})x^{\nu }=  \nonumber \\
\nonumber \\
=x^{\mu }+\frac{1}{2}\delta \omega _{\alpha \beta }(g)(I^{\alpha \beta
})_{~~\nu }^{\mu }(\left\{ x\right\} _{n.m.})x^{\nu }=  \nonumber \\
\nonumber \\
=x^{\mu }+\left( -\theta _{l}(g)S^{l}(\left\{ x\right\} _{n.m.})-\zeta
_{i}(g)K^{i}(\left\{ x\right\} _{n.m.})\right) _{~~\nu }^{\mu }x^{\nu }=
\nonumber \\
\nonumber \\
=x^{\mu }+\left( -{\bf \theta }(g)\cdot {\bf S}(\left\{ x\right\} _{n.m.})-%
{\bf \zeta }(g)\cdot {\bf K}(\left\{ x\right\} _{n.m.})\right) _{~~\nu
}^{\mu }x^{\nu }=  \nonumber \\
\nonumber \\
=\left( 1-{\bf \theta }(g)\cdot {\bf S}(\left\{ x\right\} _{n.m.})-{\bf %
\zeta }(g)\cdot {\bf K}(\left\{ x\right\} _{n.m.})\right) _{~~\nu }^{\mu
}x^{\nu }=  \nonumber \\
\nonumber \\
=(1-\theta _{1}(g)S^{1}(\left\{ x\right\} _{n.m.})-\theta
_{2}(g)S^{2}(\left\{ x\right\} _{n.m.})-\theta _{3}(g)S^{3}(\left\{
x\right\} _{n.m.})+  \nonumber \\
\nonumber \\
-\zeta _{1}(g)K^{1}(\left\{ x\right\} _{n.m.})-\zeta _{2}(g)K^{2}(\left\{
x\right\} _{n.m.})-\zeta _{3}(g)K^{3}(\left\{ x\right\} _{n.m.}))_{~~\nu
}^{\mu }x^{\nu }.
\end{gather}
Here, $1$ is the identity of $SO(S,T=4-S)_{GEN}$ , and the completely
covariant, antisymmetric 4-tensor $\delta \omega _{\alpha \beta }(g)$ is the
tensor of the 4(4-1)/2=6 dimensionless parameters of the rotation component $%
SO(S,T=4-S)_{GEN.}^{6\text{{\footnotesize \ }}}$ of the maximal Killing
group $P(S,T=4-S)_{GEN.}^{10\text{{\footnotesize \ }}}$ of the generalized
4-d. Minkowski space considered. \ As shown in [1], such a tensor can be
expressed as
\begin{equation}
\begin{array}{cc}
\delta \omega _{\alpha \beta }(g)= & \left(
\begin{array}{cccc}
0 & -\zeta ^{1}(g) & -\zeta ^{2}(g) & -\zeta ^{3}(g) \\
\zeta ^{1}(g) & 0 & -\theta ^{3}(g) & \theta ^{2}(g) \\
\zeta ^{2}(g) & \theta ^{3}(g) & 0 & -\theta ^{1}(g) \\
\zeta ^{3}(g) & -\theta ^{2}(g) & \theta ^{1}(g) & 0
\end{array}
\right) ,
\end{array}
\end{equation}
where the rotation parameter (Euclidean) axial 3-vector ${\bf \theta }(g)$
and the boost parameter (Euclidean) polar 3-vector ${\bf \zeta }(g)$ are
defined respectively by {\bf (}$\epsilon _{ijk}${\bf \ }is the Euclidean
Levi-Civita 3-tensor, with the assumed conbention{\bf \ }$\epsilon
_{123}\equiv 1${\bf )}
\begin{equation}
\theta ^{i}(g)=\theta _{i}(g)\equiv -\frac{1}{2}\epsilon _{i}^{~jk}\delta
\omega _{jk}(g),
\end{equation}
\begin{equation}
\zeta ^{i}(g)=\zeta _{i}(g)\equiv -\delta \omega _{0i}(g),
\end{equation}
namely
\begin{equation}
{\bf \theta }(g)\equiv (-\delta \omega _{23}(g),-\delta \omega
_{31}(g),-\delta \omega _{12}(g)),
\end{equation}
\begin{equation}
{\bf \zeta }(g)\equiv (-\delta \omega _{01}(g),-\delta \omega
_{02}(g),-\delta \omega _{03}(g)).
\end{equation}
Moreover, the matrix 3-vectors
\begin{equation}
{\bf S}(\left\{ x\right\} _{n.m.})\equiv \smallskip (I^{23}(\left\{
x\right\} _{n.m.}),I^{31}(\left\{ x\right\} _{n.m.}),I^{12}(\left\{
x\right\} _{n.m.})),
\end{equation}
\begin{equation}
{\bf K}(\left\{ x\right\} _{n.m.})\equiv (I^{01}(\left\{ x\right\}
_{n.m.}),I^{02}(\left\{ x\right\} _{n.m.}),I^{03}(\left\{ x\right\} _{n.m.}))%
\text{ }
\end{equation}
constitute the {\em self-representation basis }for $SO(S,T=4-S)_{GEN.}$.

From Eq. (9) it follows that the finite space-time rotation in $\widetilde{%
M_{4}}(\left\{ x\right\} _{n.m.})$ corresponding to $g\in SO(S,T=4-S)_{GEN.}$
can be written as:
\begin{gather}
SO(S,T=4-S)_{GEN.}\ni g:  \nonumber \\
\nonumber \\
x^{\mu }\rightarrow \left( x^{\prime }\right) _{(g)}^{\mu }(\left\{
x\right\} _{m.},\left\{ x\right\} _{n.m.})=\left( x^{\mu }\right)
_{(g)}^{^{\prime }}(\left\{ x\right\} _{m.},\left\{ x\right\} _{n.m.})=
\nonumber \\
\nonumber \\
=\exp (-{\bf \theta }(g)\cdot {\bf S}(\left\{ x\right\} _{n.m.})-{\bf \zeta }%
(g)\cdot {\bf K}(\left\{ x\right\} _{n.m.}))_{~~\nu }^{\mu }x^{\nu }=
\nonumber \\
\nonumber \\
=\exp (-\theta _{1}(g)S^{1}(\left\{ x\right\} _{n.m.})-\theta
_{2}(g)S^{2}(\left\{ x\right\} _{n.m.})-\theta _{3}(g)S^{3}(\left\{
x\right\} _{n.m.})+  \nonumber \\
\nonumber \\
-\zeta _{1}(g)K^{1}(\left\{ x\right\} _{n.m.})-\zeta _{2}(g)K^{2}(\left\{
x\right\} _{n.m.})-\zeta _{3}(g)K^{3}(\left\{ x\right\} _{n.m.}))_{~~\nu
}^{\mu }x^{\nu }\smallskip ,
\end{gather}
or, by a series development of the exponential:
\begin{gather}
SO(S,T=4-S)_{GEN.}\ni g:  \nonumber \\
\nonumber \\
x^{\mu }\rightarrow \left( x^{\prime }\right) _{(g)}^{\mu }(\left\{
x\right\} _{m.},\left\{ x\right\} _{n.m.})=\left( x^{\mu }\right)
_{(g)}^{^{\prime }}(\left\{ x\right\} _{m.},\left\{ x\right\} _{n.m.})=
\nonumber \\
\nonumber \\
=\left( \sum_{n=0}^{\infty }\frac{1}{n!}(-{\bf \theta }(g)\cdot {\bf S}%
(\left\{ x\right\} _{n.m.})-{\bf \zeta }(g)\cdot {\bf K}(\left\{ x\right\}
_{n.m.}))^{n}\right) _{~~\nu }^{\mu }x^{\nu }=  \nonumber \\
\nonumber \\
=(\sum_{n=0}^{\infty }\frac{1}{n!}(-\theta _{1}(g)S^{1}(\left\{ x\right\}
_{n.m.})-\theta _{2}(g)S^{2}(\left\{ x\right\} _{n.m.})-\theta
_{3}(g)S^{3}(\left\{ x\right\} _{n.m.})+  \nonumber \\
\nonumber \\
-\zeta _{1}(g)K^{1}(\left\{ x\right\} _{n.m.})-\zeta _{2}(g)K^{2}(\left\{
x\right\} _{n.m.})-\zeta _{3}(g)K^{3}(\left\{ x\right\} _{n.m.}))^{n})_{~\nu
}^{\mu }x^{\nu }\smallskip .
\end{gather}
Since the generalized Lorentz algebra (8) is non-commutative, the group $%
SO(S,T)_{GEN.}^{N(N-1)/2\text{{\small \ }}}$ \ is non-Abelian, and therefore
the 4-d. finite transformations (17), (18) do not commute.

\subsection{Finite\ Transformations\ of the\ Deformed, Homogeneous\ Lorentz
Group $SO(3,1)_{DEF.}$ of DSR4.}

In the case of the deformed Minkowski space-time $\widetilde{M_{4}}(x^{5})$
(corresponding to $N=4$, $S=3$, $T=1$, $x^{\mu }=\left\{
x^{0},x^{1},x^{2},x^{3}\right\} $, $\left\{ x\right\} _{n.m.}=x^{5}$) the
{\em generalized Lorentz algebra }(8){\em \ }becomes the {\em (4-d.)
deformed Lorentz algebra} $su(2)_{DEF.}\times su(2)_{DEF.}$, and the
infinitesimal space-time rotation (9) reads
\begin{gather}
su(2)_{DEF.}\times su(2)_{DEF.}\ni \delta g:  \nonumber \\
\nonumber \\
x^{\mu }\rightarrow \left( x^{\prime }\right) _{(g),DSR4}^{\mu }(\left\{
x\right\} _{m.},x^{5})=\left( x^{\mu }\right) _{(g),DSR4}^{^{\prime
}}(\left\{ x\right\} _{m.},x^{5})=  \nonumber \\
\nonumber \\
=x^{\mu }+\delta x_{(g),DSR4}^{\mu }(\left\{ x\right\} _{m.},x^{5})=
\nonumber \\
\nonumber \\
=x^{\mu }+\left( -\theta _{l}(g)S_{DSR4}^{l}(x^{5})-\zeta
_{i}(g)K_{DSR4}^{i}(x^{5})\right) _{~~\nu }^{\mu }x^{\nu }=  \nonumber \\
\nonumber \\
=x^{\mu }+\left( -{\bf \theta }(g)\cdot {\bf S}_{DSR4}(x^{5})-{\bf \zeta }%
(g)\cdot {\bf K}_{DSR4}(x^{5})\right) _{~~\nu }^{\mu }x^{\nu }=  \nonumber \\
\nonumber \\
=\left( 1-{\bf \theta }(g)\cdot {\bf S}_{DSR4}(x^{5})-{\bf \zeta }(g)\cdot
{\bf K}_{DSR4}(x^{5})\right) _{~~\nu }^{\mu }x^{\nu }=  \nonumber \\
\nonumber \\
=(1-\theta _{1}(g)S_{DSR4}^{1}(x^{5})-\theta
_{2}(g)S_{DSR4}^{2}(x^{5})-\theta _{3}(g)S_{DSR4}^{3}(x^{5})+  \nonumber \\
\nonumber \\
-\zeta _{1}(g)K_{DSR4}^{1}(x^{5})-\zeta _{2}(g)K_{DSR4}^{2}(x^{5})-\zeta
_{3}(g)K_{DSR4}^{3}(x^{5}))_{~~\nu }^{\mu }x^{\nu },
\end{gather}
with obvious meaning of notation (in particular $1$ denotes the identity of $%
SO(3,1)_{DEF.}$, corresponding -by the homomorphic exponential mapping- to
the origin of the algebra $su(2)_{DEF.}\times su(2)_{DEF.}$).

In this case (like in the standard SR), since the couple $\left( S,T\right)
=\left( 3,1\right) $ denoting the hyperbolic metric signature of $\widetilde{%
M_{4}}(x^{5})$ is fixed, it is possible to give a physical interpretation of
parameters and generators: ${\bf \theta }(g)$ and ${\bf \zeta }(g)$ are,
respectively, the deformed rotation and boost Euclidean 3-vectors, whereas $%
{\bf S}_{DSR4}(x^{5})$ and ${\bf K}_{DSR4}(x^{5})$ are the (operatorial
3-vectors of) generators of the corresponding transformations of the
deformed, homogeneous Lorentz group $SO(3,1)_{DEF.}$, satisfying the {\em %
(4-d.) deformed Lorentz algebra} $su(2)_{DEF.}\times su(2)_{DEF.}$ :
\begin{equation}
\left\{
\begin{array}{l}
\lbrack S_{DSR4}^{i}(x^{5}),S_{DSR4}^{j}(x^{5})]=\epsilon
_{ijk}b_{k}^{-2}(x^{5})S_{DSR4}^{k}(x^{5})\medskip ; \\
\\
\lbrack K_{DSR4}^{i}(x^{5}),K_{DSR4}^{j}(x^{5})]=-b_{0}^{-2}(x^{5})\epsilon
_{ijk}S_{DSR4}^{k}(x^{5})\medskip ; \\
\\
\lbrack S_{DSR4}^{i}(x^{5}),K_{DSR4}^{j}(x^{5})]=\epsilon
_{ijl}b_{j}^{-2}K_{DSR4}^{l}(x^{5})\medskip .
\end{array}
\right.
\end{equation}
In the DSR4 case, Eq. (18) for a finite transformation becomes

\begin{gather}
SO(3,1)_{DEF.}\ni g:  \nonumber \\
\nonumber \\
x^{\mu }\rightarrow \left( x^{\prime }\right) _{(g),DSR4}^{\mu }(\left\{
x\right\} _{m.},x^{5})=\left( x^{\mu }\right) _{(g),DSR4}^{^{\prime
}}(\left\{ x\right\} _{m.},x^{5})=  \nonumber \\
\nonumber \\
=\exp (-{\bf \theta }(g)\cdot {\bf S}_{DSR4}(x^{5})-{\bf \zeta }(g)\cdot
{\bf K}_{DSR4}(x^{5}))_{~~\nu }^{\mu }x^{\nu }=  \nonumber \\
\nonumber \\
=\exp (-\theta _{1}(g)S_{DSR4}^{1}(x^{5})-\theta
_{2}(g)S_{DSR4}^{2}(x^{5})-\theta _{3}(g)S_{DSR4}^{3}(x^{5})+  \nonumber \\
\nonumber \\
-\zeta _{1}(g)K_{DSR4}^{1}(x^{5})-\zeta _{2}(g)K_{DSR4}^{2}(x^{5})-\zeta
_{3}(g)K_{DSR4}^{3}(x^{5}))_{~~\nu }^{\mu }x^{\nu }=  \nonumber \\
\nonumber \\
=\left( \sum_{n=0}^{\infty }\frac{1}{n!}(-{\bf \theta }(g)\cdot {\bf S}%
_{DSR4}(x^{5})-{\bf \zeta }(g)\cdot {\bf K}_{DSR4}(x^{5})))^{n}\right)
_{~~\nu }^{\mu }x^{\nu }=  \nonumber \\
\nonumber \\
=(\sum_{n=0}^{\infty }\frac{1}{n!}(-\theta _{1}(g)S_{DSR4}^{1}(x^{5})-\theta
_{2}(g)S_{DSR4}^{2}(x^{5})-\theta _{3}(g)S_{DSR4}^{3}(x^{5})+  \nonumber \\
\nonumber \\
-\zeta _{1}(g)K_{DSR4}^{1}(x^{5})-\zeta _{2}(g)K_{DSR4}^{2}(x^{5})-\zeta
_{3}(g)K_{DSR4}^{3}(x^{5}))^{n})_{~~\nu }^{\mu }x^{\nu }\smallskip .
\end{gather}
\pagebreak

\section{FINITE\ SPACE-TIME\ ROTATIONS ABOUT\ A\ COORDINATE\ AXIS IN\ $%
\widetilde{M_{4}}(x^{5})$}

\subsection{Infinitesimal Generators\protect\bigskip}

Let us recall the explicit form of the matrices of the 4-d. representation
of the infinitesimal generators of the group $SO(3,1)_{DEF.}$, derived in
[1]:
\begin{equation}
\begin{array}{cc}
I_{DSR4}^{10}(x^{5})= & \left(
\begin{array}{cccc}
0 & -b_{0}^{-2}(x^{5}) & 0 & 0 \\
-b_{1}^{-2}(x^{5}) & 0 & 0 & 0 \\
0 & 0 & 0 & 0 \\
0 & 0 & 0 & 0
\end{array}
\right) ;
\end{array}
\end{equation}
\begin{equation}
\begin{array}{cc}
I_{DSR4}^{20}(x^{5})= & \left(
\begin{array}{cccc}
0 & 0 & -b_{0}^{-2}(x^{5}) & 0 \\
0 & 0 & 0 & 0 \\
-b_{2}^{-2}(x^{5}) & 0 & 0 & 0 \\
0 & 0 & 0 & 0
\end{array}
\right) ;
\end{array}
\end{equation}
\begin{equation}
\begin{array}{cc}
I_{DSR4}^{30}(x^{5})= & \left(
\begin{array}{cccc}
0 & 0 & 0 & -b_{0}^{-2}(x^{5}) \\
0 & 0 & 0 & 0 \\
0 & 0 & 0 & 0 \\
-b_{3}^{-2}(x^{5}) & 0 & 0 & 0
\end{array}
\right) ;
\end{array}
\end{equation}
\begin{equation}
\begin{array}{cc}
I_{DSR4}^{12}(x^{5})= & \left(
\begin{array}{cccc}
0 & 0 & 0 & 0 \\
0 & 0 & -b_{1}^{-2}(x^{5}) & 0 \\
0 & b_{2}^{-2}(x^{5}) & 0 & 0 \\
0 & 0 & 0 & 0
\end{array}
\right) ;
\end{array}
\end{equation}
\begin{equation}
\begin{array}{cc}
I_{DSR4}^{23}(x^{5})= & \left(
\begin{array}{cccc}
0 & 0 & 0 & 0 \\
0 & 0 & 0 & 0 \\
0 & 0 & 0 & -b_{2}^{-2}(x^{5}) \\
0 & 0 & b_{3}^{-2}(x^{5}) & 0
\end{array}
\right) ;
\end{array}
\end{equation}
\begin{equation}
\begin{array}{cc}
I_{DSR4}^{31}(x^{5})= & \left(
\begin{array}{cccc}
0 & 0 & 0 & 0 \\
0 & 0 & 0 & b_{1}^{-2}(x^{5}) \\
0 & 0 & 0 & 0 \\
0 & -b_{3}^{-2}(x^{5}) & 0 & 0
\end{array}
\right) .
\end{array}
\end{equation}

In the following, we shall need the (not maximally-ranked) 4$\times $4
matrices $A_{i}$ , $B_{i}$ ($i=1,2,3$), defined by:
\begin{equation}
A_{1}\equiv \left(
\begin{array}{cccc}
1 & 0 & 0 & 0 \\
0 & 1 & 0 & 0 \\
0 & 0 & 0 & 0 \\
0 & 0 & 0 & 0
\end{array}
\right) ;
\end{equation}
\begin{equation}
A_{2}\equiv \left(
\begin{array}{cccc}
1 & 0 & 0 & 0 \\
0 & 0 & 0 & 0 \\
0 & 0 & 1 & 0 \\
0 & 0 & 0 & 0
\end{array}
\right) ;
\end{equation}
\begin{equation}
A_{3}\equiv \left(
\begin{array}{cccc}
1 & 0 & 0 & 0 \\
0 & 0 & 0 & 0 \\
0 & 0 & 0 & 0 \\
0 & 0 & 0 & 1
\end{array}
\right) ;
\end{equation}

\begin{equation}
B_{1}\equiv \left(
\begin{array}{cccc}
0 & 0 & 0 & 0 \\
0 & 0 & 0 & 0 \\
0 & 0 & 1 & 0 \\
0 & 0 & 0 & 1
\end{array}
\right) ;
\end{equation}
\begin{equation}
B_{2}\equiv \left(
\begin{array}{cccc}
0 & 0 & 0 & 0 \\
0 & 1 & 0 & 0 \\
0 & 0 & 0 & 0 \\
0 & 0 & 0 & 1
\end{array}
\right) ;
\end{equation}
\begin{equation}
B_{3}\equiv \left(
\begin{array}{cccc}
0 & 0 & 0 & 0 \\
0 & 1 & 0 & 0 \\
0 & 0 & 1 & 0 \\
0 & 0 & 0 & 0
\end{array}
\right) .
\end{equation}

It is easy to see, by explicit calculation, that the powers of the
generators of deformed boosts and rotations satisfy, respectively, the
relations (ESC off) :
\begin{equation}
\left\{
\begin{array}{l}
\left( I_{DSR4}^{i0}(x^{5})\right) ^{0}=1_{4-d.}; \\
\\
\left( I_{DSR4}^{i0}(x^{5})\right)
^{2n}=b_{0}^{-2n}(x^{5})b_{i}^{-2n}(x^{5})A_{i}\text{ },n\in N; \\
\\
\left( I_{DSR4}^{i0}(x^{5})\right)
^{2n+1}=b_{0}^{-2n}(x^{5})b_{i}^{-2n}(x^{5})I_{DSR4}^{i0}(x^{5})\text{ }%
,n\in N\cup \left\{ 0\right\} \medskip ;
\end{array}
\right.
\end{equation}

\begin{equation}
\left\{
\begin{array}{l}
\left( I_{DSR4}^{ij}(x^{5})\right) ^{0}=1_{4-d.}; \\
\\
\\
\left( I_{DSR4}^{ij}(x^{5})\right)
^{2n}=(-1)^{n}b_{i}^{-2n}(x^{5})b_{j}^{-2n}(x^{5})B_{s\neq i,s\neq j}= \\
\\
=(-1)^{n}b_{i}^{-2n}(x^{5})b_{j}^{-2n}(x^{5})\left( \sum_{s=1}^{3}(1-\delta
_{si})(1-\delta _{sj})B_{s}\right) = \\
\\
\stackrel{\text{ESC off on{\footnotesize \ }}i\text{{\footnotesize \ }and%
{\footnotesize \ }}j\text{, ESC on on }k}{=}\left( -1\right)
^{n}b_{i}^{-2n}(x^{5})b_{j}^{-2n}(x^{5})\left| \epsilon _{ijk}\right| B_{k}%
\text{ },n\in N\medskip \text{ }; \\
\\
\\
\left( I_{DSR4}^{ij}(x^{5})\right)
^{2n+1}=(-1)^{n}b_{i}^{-2n}(x^{5})b_{j}^{-2n}(x^{5})I_{DSR4}^{ij}(x^{5})%
\text{ },n\in N\cup \left\{ 0\right\} \medskip ,
\end{array}
\right.
\end{equation}
($i=1,2,3$, with $1_{4-d.}$ being the identity 4$\times $4 matrix and $%
\left| \epsilon _{ijk}\right| \equiv sgn\left( \epsilon _{ijk}\right)
\epsilon _{ijk}$). Notice that Eqs. (34) and (35), although obtained by
utilizing a 4-d. representation of the infinitesimal generators of $%
SO(3,1)_{DEF.}$, hold true in general at abstract group level (i.e. they are
{\em representation-independent}).

\subsection{Finite Deformed Boost Along a Coordinate Axis}

Let us first consider a finite deformed boost transformation, with rapidity
parameter $\zeta _{1}(g)$ along $\widehat{x^{1}}$. Recalling that $%
K_{DSR4}^{i}(x^{5})\equiv $ $I_{DSR4}^{0i}(x^{5})$ $\forall i=1,2,3$, we
get, from Eq. (21):\bigskip\
\begin{gather}
SO(3,1)_{DEF.}\ni g:  \nonumber \\
\nonumber \\
x^{\mu }\rightarrow \left( x^{\prime }\right) _{(g),DSR4}^{\mu }(\left\{
x\right\} _{m.},x^{5})=\left( x^{\mu }\right) _{(g),DSR4}^{^{\prime
}}(\left\{ x\right\} _{m.},x^{5})=  \nonumber \\
\nonumber \\
=\exp (-\zeta _{1}(g)K_{DSR4}^{1}(x^{5}))_{~~\nu }^{\mu }x^{\nu }=\exp
(\zeta _{1}(g)I_{DSR4}^{10}(x^{5}))_{~~\nu }^{\mu }x^{\nu }=  \nonumber \\
\nonumber \\
=\left( \sum_{n=0}^{\infty }\frac{1}{n!}(\zeta
_{1}(g)I_{DSR4}^{10}(x^{5}))^{n}\right) _{~~\nu }^{\mu }x^{\nu }=\left(
\sum_{n=0}^{\infty }\frac{1}{n!}(\zeta
_{1}(g))^{n}(I_{DSR4}^{10}(x^{5}))^{n}\right) _{~~\nu }^{\mu }x^{\nu }=
\nonumber \\
\nonumber \\
=\left( 1_{4-d.}+\sum_{n=1}^{\infty }\frac{1}{\left( 2n\right) !}(\zeta
_{1}(g))^{2n}(I_{DSR4}^{10}(x^{5}))^{2n}+\right.  \nonumber \\
\nonumber \\
\left. +\sum_{n=0}^{\infty }\frac{1}{\left( 2n+1\right) !}(\zeta
_{1}(g))^{2n+1}(I_{DSR4}^{10}(x^{5}))_{~~}^{2n+1}\right) _{~~\nu }^{\mu
}x^{\nu }=  \nonumber \\
\nonumber \\
=\left( 1_{4-d.}+A_{1}\sum_{n=1}^{\infty }\frac{1}{\left( 2n\right) !}%
b_{0}^{-2n}(x^{5})b_{1}^{-2n}(x^{5})(\zeta _{1}(g))^{2n}\right. +  \nonumber
\\
\nonumber \\
\left. +I_{DSR4}^{10}(x^{5})\sum_{n=0}^{\infty }\frac{1}{\left( 2n+1\right) !%
}b_{0}^{-2n}(x^{5})b_{1}^{-2n}(x^{5})(\zeta _{1}(g))^{2n+1}\right) _{~~\nu
}^{\mu }x^{\nu }=  \nonumber \\
\nonumber \\
=\left( 1_{4-d.}+A_{1}\sum_{n=1}^{\infty }\frac{1}{\left( 2n\right) !}(\zeta
_{1}(g)b_{0}^{-1}(x^{5})b_{1}^{-1}(x^{5}))^{2n}+\right.  \nonumber \\
\nonumber \\
\left. +I_{DSR4}^{10}(x^{5})b_{0}(x^{5})b_{1}(x^{5})\sum_{n=0}^{\infty }%
\frac{1}{\left( 2n+1\right) !}(\zeta
_{1}(g)b_{0}^{-1}(x^{5})b_{1}^{-1}(x^{5}))^{2n+1}\right) _{~~\nu }^{\mu
}x^{\nu }=  \nonumber \\
\nonumber \\
=\left( 1_{4-d.}+\left( \left( \cosh \zeta
_{1}(g)b_{0}^{-1}(x^{5})b_{1}^{-1}(x^{5})\right) -1\right) A_{1}+\right.
\nonumber \\
\nonumber \\
\left. +b_{0}(x^{5})b_{1}(x^{5})\left( \sinh \zeta
_{1}(g)b_{0}^{-1}(x^{5})b_{1}^{-1}(x^{5})\right) I_{DSR4~\
}^{10}(x^{5})\right) _{~~\nu }^{\mu }x\smallskip ^{\nu },
\end{gather}
where in the last passage the series expansions of hyperbolic functions have
been used.

By defining the 4$\times $4 matrix
\begin{gather}
Boost_{DSR4,\widehat{x^{1}}}(g,x^{5})\equiv 1_{4-d.}+\left( \left( \cosh
\zeta _{1}(g)b_{0}^{-1}(x^{5})b_{1}^{-1}(x^{5})\right) -1\right) A_{1}+
\nonumber \\
\nonumber \\
+b_{0}(x^{5})b_{1}(x^{5})\left( \sinh \zeta
_{1}(g)b_{0}^{-1}(x^{5})b_{1}^{-1}(x^{5})\right) I_{DSR4~\
}^{10}(x^{5})\smallskip ,
\end{gather}
Eq. (36) can be rewritten as:
\begin{gather}
SO(3,1)_{DEF.}\ni g:x^{\mu }\rightarrow \left( x^{\prime }\right)
_{(g),DSR4}^{\mu }(\left\{ x\right\} _{m.},x^{5})=\left( x^{\mu }\right)
_{(g),DSR4}^{^{\prime }}(\left\{ x\right\} _{m.},x^{5})=  \nonumber \\
\nonumber \\
=\left( Boost_{DSR4,\widehat{x^{1}}}(g,x^{5})\smallskip \right) _{~~\nu
}^{\mu }x^{\nu }.
\end{gather}

We find therefore
\begin{gather}
\left(
\begin{array}{c}
\left( x^{\prime }\right) _{(g),DSR4}^{0}(\left\{ x\right\} _{m.},x^{5}) \\
\left( x^{\prime }\right) _{(g),DSR4}^{1}(\left\{ x\right\} _{m.},x^{5}) \\
\left( x^{\prime }\right) _{(g),DSR4}^{2}(\left\{ x\right\} _{m.},x^{5}) \\
\left( x^{\prime }\right) _{(g),DSR4}^{3}(\left\{ x\right\} _{m.},x^{5})
\end{array}
\right) =\left(
\begin{array}{c}
\left( x^{0}\right) _{(g),DSR4}^{^{\prime }}(\left\{ x\right\} _{m.},x^{5})
\\
\left( x^{1}\right) _{(g),DSR4}^{^{\prime }}(\left\{ x\right\} _{m.},x^{5})
\\
\left( x^{2}\right) _{(g),DSR4}^{^{\prime }}(\left\{ x\right\} _{m.},x^{5})
\\
\left( x^{3}\right) _{(g),DSR4}^{^{\prime }}(\left\{ x\right\} _{m.},x^{5})
\end{array}
\right) =  \nonumber \\
\nonumber \\
=\left( Boost_{DSR4,\widehat{x^{1}}}(g,x^{5})\right) \times \left(
\begin{array}{c}
x^{0} \\
x^{1} \\
x^{2} \\
x^{3}
\end{array}
\right) \smallskip =  \nonumber \\
\\
=\left(
\begin{array}{c}
\left( \cosh \zeta _{1}(g)b_{0}^{-1}(x^{5})b_{1}^{-1}(x^{5})\right)
x^{0}-b_{0}^{-1}(x^{5})b_{1}(x^{5})\left( \sinh \zeta
_{1}(g)b_{0}^{-1}(x^{5})b_{1}^{-1}(x^{5})\right) x^{1} \\
\\
-b_{0}(x^{5})b_{1}^{-1}(x^{5})\left( \sinh \zeta
_{1}(g)b_{0}^{-1}(x^{5})b_{1}^{-1}(x^{5})\right) x^{0}+\left( \cosh \zeta
_{1}(g)b_{0}^{-1}(x^{5})b_{1}^{-1}(x^{5})\right) x^{1} \\
\\
x^{2} \\
\\
x^{3}
\end{array}
\right) \smallskip .  \nonumber
\end{gather}
Analogous results are obtained for the finite deformed boosts along the
other two spatial axes. By defining the matrices
\begin{gather}
Boost_{DSR4,\widehat{x^{2}}}(g,x^{5})\equiv 1_{4-d.}+\left( \left( \cosh
\zeta _{2}(g)b_{0}^{-1}(x^{5})b_{2}^{-1}(x^{5})\right) -1\right)
A_{2}+\smallskip  \nonumber \\
\nonumber \\
+b_{0}(x^{5})b_{2}(x^{5})\left( \sinh \zeta
_{2}(g)b_{0}^{-1}(x^{5})b_{2}^{-1}(x^{5})\right) I_{DSR4~\
}^{20}(x^{5})\smallskip ;  \nonumber \\
\end{gather}
\begin{gather}
Boost_{DSR4,\widehat{x^{3}}}(g,x^{5})\equiv 1_{4-d.}+\left( \left( \cosh
\zeta _{3}(g)b_{0}^{-1}(x^{5})b_{3}^{-1}(x^{5})\right) -1\right) A_{3}+
\nonumber \\
\nonumber \\
+b_{0}(x^{5})b_{3}(x^{5})\left( \sinh \zeta
_{3}(g)b_{0}^{-1}(x^{5})b_{3}^{-1}(x^{5})\right) I_{DSR4~\
}^{30}(x^{5})\smallskip ,  \nonumber \\
\end{gather}
we get indeed
\begin{gather}
SO(3,1)_{DEF.}\ni g:x^{\mu }\rightarrow \left( x^{\prime }\right)
_{(g),DSR4}^{\mu }(\left\{ x\right\} _{m.},x^{5})=\left( x^{\mu }\right)
_{(g),DSR4}^{^{\prime }}(\left\{ x\right\} _{m.},x^{5})=  \nonumber \\
\nonumber \\
=\left( Boost_{DSR4,\widehat{x^{2}}}(g,x^{5})\smallskip \right) _{~~\nu
}^{\mu }x^{\nu };  \nonumber \\
\end{gather}
\begin{gather}
SO(3,1)_{DEF.}\ni g:x^{\mu }\rightarrow \left( x^{\prime }\right)
_{(g),DSR4}^{\mu }(\left\{ x\right\} _{m.},x^{5})=\left( x^{\mu }\right)
_{(g),DSR4}^{^{\prime }}(\left\{ x\right\} _{m.},x^{5})=  \nonumber \\
\nonumber \\
=\left( Boost_{DSR4,\widehat{x^{3}}}(g,x^{5})\smallskip \right) _{~~\nu
}^{\mu }x^{\nu }.  \nonumber \\
\end{gather}
\bigskip

Let us introduce the {\em (effective) deformed rapidity} ${\bf \tilde{\zeta}}%
(g,x^{5})$, defined by\bigskip\
\begin{equation}
\tilde{\zeta}_{i}(g,x^{5})\equiv \zeta
_{i}(g)b_{0}^{-1}(x^{5})b_{i}^{-1}(x^{5})\text{ \ }\forall i=1,2,3.
\end{equation}
\bigskip Then, a finite deformed boost transformation, with rapidity
parameter $\zeta _{i}(g)$ along $\widehat{x^{i}}$ can be written in compact
form as
\begin{gather*}
\left( x^{\prime }\right) _{(g),DSR4}^{0}(\left\{ x\right\}
_{m.},x^{5})=\left( x^{0}\right) _{(g),DSR4}^{^{\prime }}(\left\{ x\right\}
_{m.},x^{5})= \\
\\
=\left( \cosh \tilde{\zeta}_{i}(g,x^{5})\right)
x^{0}-b_{0}^{-1}(x^{5})b_{i}(x^{5})\left( \sinh \tilde{\zeta}%
_{i}(g,x^{5})\right) x^{i}, \\
\\
\end{gather*}
\begin{gather*}
\left( x^{\prime }\right) _{(g),DSR4}^{i}(\left\{ x\right\}
_{m.},x^{5})=\left( x^{i}\right) _{(g),DSR4}^{^{\prime }}(\left\{ x\right\}
_{m.},x^{5})= \\
\\
=-b_{0}(x^{5})b_{i}^{-1}(x^{5})\left( \sinh \tilde{\zeta}_{i}(g,x^{5})%
\right) x^{0}+\left( \cosh \tilde{\zeta}_{i}(g,x^{5})\right) x^{i}, \\
\\
\end{gather*}
\begin{equation}
\left( x^{\prime }\right) _{(g),DSR4}^{k\neq i}(\left\{ x\right\}
_{m.},x^{5})=\left( x^{k\neq i}\right) _{(g),DSR4}^{^{\prime }}(\left\{
x\right\} _{m.},x^{5})=x^{k\neq i}\bigskip .
\end{equation}

\bigskip By recalling the expression of a boost in SR(4)
\begin{eqnarray}
&&\left\{
\begin{array}{l}
\left( x^{\prime }\right) _{(g),SR4}^{0}(\left\{ x\right\} _{m.})=\left(
x^{0}\right) _{(g),SR4}^{^{\prime }}(\left\{ x\right\} _{m.})=\left( \cosh
\zeta _{i}(g)\right) x^{0}-\left( \sinh \zeta _{i}(g)\right) x^{i} \\
\\
\left( x^{\prime }\right) _{(g),SR4}^{i}(\left\{ x\right\} _{m.})=\left(
x^{i}\right) _{(g),SR4}^{^{\prime }}(\left\{ x\right\} _{m.})=-\left( \sinh
\zeta _{i}(g)\right) x^{0}+\left( \cosh \zeta _{i}(g)\right) x^{i}\medskip
\\
\\
\left( x^{\prime }\right) _{(g),SR4}^{k\neq i}(\left\{ x\right\}
_{m.})=\left( x^{k\neq i}\right) _{(g),SR4}^{^{\prime }}(\left\{ x\right\}
_{m.})=x^{k\neq i}\medskip ,
\end{array}
\right.  \nonumber \\
&&
\end{eqnarray}
it is easily seen that the deforming transition SR$\rightarrow $DSR4
corresponds --- at the level of group parameters ---to the {\em deforming
and anisotropizing rescaling} of rapidities $\zeta _{i}(g)\rightarrow \tilde{%
\zeta}_{i}(g,x^{5})$, $\forall i=1,2,3$.

\subsubsection{Parametric Change of Basis for a Deformed Boost along a
Coordinate Axis.}

We recall that a deformed boost with speed parameter $v^{i}$ along $\widehat{%
x^{i}}$ reads (see Ref. [10]) (ESC off throughout):
\begin{gather*}
\left( x^{\prime }\right) _{(g),DSR4}^{i}(\left\{ x\right\}
_{m.},x^{5})=\left( x^{i}\right) _{(g),DSR4}^{^{\prime }}(\left\{ x\right\}
_{m.},x^{5})= \\
\\
=\widetilde{\gamma }(g)(x^{i}-v^{i}(g)t)=\ \ \widetilde{\gamma }(g)\left(
x^{i}-\widetilde{\beta }(g)\dfrac{b_{0}(x^{5})}{b_{i}(x^{5})}ct\right) , \\
\\
\end{gather*}
\begin{eqnarray*}
\left( x^{\prime }\right) _{(g),DSR4}^{k\neq i}(\left\{ x\right\}
_{m.},x^{5}) &=&\left( x^{k\neq i}\right) _{(g),DSR4}^{^{\prime }}(\left\{
x\right\} _{m.},x^{5})=x^{k\neq i}, \\
&& \\
&&
\end{eqnarray*}
\begin{gather}
t_{(g),DSR4}^{\prime }(\left\{ x\right\} _{m.},x^{5})=  \nonumber \\
\nonumber \\
=\widetilde{\gamma }(g)\left( t-\dfrac{v^{i}(g)b_{i}^{2}(x^{5})}{%
c^{2}b_{0}^{2}(x^{5})}x^{i}\right) =\widetilde{\gamma }(g)\left( t-\dfrac{%
\widetilde{\beta }^{2}(g)}{v^{i}(g)}x^{i}\right) ,
\end{gather}
where (the dependence on $x^{5}$ is omitted, but understood)

\begin{equation}
\widetilde{\beta }(g)=\widetilde{\beta ^{i}}(g)\equiv \frac{%
v^{i}(g)b_{i}(x^{5})}{cb_{0}(x^{5})}\equiv \frac{v^{i}(g)}{u_{i}};
\end{equation}
\begin{equation}
\widetilde{\gamma }(g)=\widetilde{\gamma ^{i}}(g)\equiv \left( 1-\left(
\widetilde{\beta ^{i}}(g)\right) ^{2}\right) ^{-1/2}=\left( 1-\left( \frac{%
v^{i}(g)b_{i}(x^{5})}{cb_{0}(x^{5})}\right) ^{2}\right) ^{-1/2}\smallskip .
\end{equation}
Quantity $u_{i}$ is the maximal causal speed (along $\widehat{x^{i}}$ ) in $%
\widetilde{M_{4}}(x_{5})$, i.e. the generalization of the speed of light for
SR ([2], [10]). In general, we have
\begin{equation}
{\bf u}\equiv \left( c\frac{b_{0}(x^{5})}{b_{1}(x^{5})}\widehat{x},c\frac{%
b_{0}(x^{5})}{b_{2}(x^{5})}\widehat{y},c\frac{b_{0}(x^{5})}{b_{3}(x^{5})}%
\widehat{z}\right) .
\end{equation}

Eq. (47) can be put in symmetrical form with respect to time and space
coordinates by introducing the dimensional coordinate $\tilde{x}^{0}$
defined by [10]
\begin{equation}
\tilde{x}^{0}\equiv u^{i}t=c\frac{b_{0}(x^{5})}{b_{i}(x^{5})}t.
\end{equation}
One gets
\begin{eqnarray}
&&\left\{
\begin{array}{l}
\left( x^{\prime }\right) _{(g),DSR4}^{i}(\left\{ x\right\}
_{m.},x^{5})=\left( x^{i}\right) _{(g),DSR4}^{^{\prime }}(\left\{ x\right\}
_{m.},x^{5})=\widetilde{\gamma }(g)(x^{i}-\widetilde{\beta ^{i}}(g)\tilde{x}%
^{0}) \\
\\
\left( x^{\prime }\right) _{(g),DSR4}^{k\neq i}(\left\{ x\right\}
_{m.},x^{5})=\left( x^{k\neq i}\right) _{(g),DSR4}^{^{\prime }}(\left\{
x\right\} _{m.},x^{5})=x^{k\neq i} \\
\\
\left( \tilde{x}^{0}\right) _{(g),DSR4}^{\prime }(\left\{ x\right\}
_{m.},x^{5})=\widetilde{\gamma }(g)(\tilde{x}^{0}-\widetilde{\beta ^{i}}%
(g)x^{i}).
\end{array}
\right.  \nonumber \\
&&
\end{eqnarray}
Such a symmetry is lost if we use the ''standard'' time coordinate $%
x^{0}\equiv ct$, which is related to $\tilde{x}^{0}$ by

\begin{equation}
\tilde{x}^{0}=x^{0}\frac{b_{0}(x^{5})}{b_{i}(x^{5})}.
\end{equation}
In terms of $x^{0}$, we have in fact
\begin{eqnarray}
&&\left\{
\begin{array}{l}
\left( x^{\prime }\right) _{(g),DSR4}^{i}(\left\{ x\right\}
_{m.},x^{5})=\left( x^{i}\right) _{(g),DSR4}^{^{\prime }}(\left\{ x\right\}
_{m.},x^{5})=\widetilde{\gamma }(g)\left( x^{i}-\widetilde{\beta ^{i}}(g)%
\dfrac{b_{0}(x^{5})}{b_{i}(x^{5})}x^{0}\bigskip \right) \\
\\
\left( x^{\prime }\right) _{(g),DSR4}^{k\neq i}(\left\{ x\right\}
_{m.},x^{5})=\left( x^{k\neq i}\right) _{(g),DSR4}^{^{\prime }}(\left\{
x\right\} _{m.},x^{5})=x^{k\neq i} \\
\\
\left( x^{\prime }\right) _{(g),DSR4}^{0}(\left\{ x\right\}
_{m.},x^{5})=\left( x^{0}\right) _{(g),DSR4}^{^{\prime }}(\left\{ x\right\}
_{m.},x^{5})=\widetilde{\gamma }(g)\left( x^{0}-\widetilde{\beta ^{i}}(g)%
\dfrac{b_{i}(x^{5})}{b_{0}(x^{5})}x^{i}\right) .
\end{array}
\right.  \nonumber \\
&&
\end{eqnarray}

Comparing Eq. (54) with Eq. (45) allows us to get the relations connecting
the dimensional parametric basis of velocities $\left\{ v^{i}\right\} $
(introduced in Ref. [10]) and the dimensionless basis of (effective)
deformed rapidities $\left\{ \widetilde{\zeta }^{i}(g,x^{5})\right\} $
(defined by (44)) (the dependence on $x^{5}$ is now fully
explicited):\bigskip\
\begin{gather}
\nonumber \\
\forall i=1,2,3\left\{
\begin{array}{l}
I)\text{ \ }\cosh \tilde{\zeta}_{i}(g,x^{5})=\widetilde{\gamma }(g,x^{5})=
\\
\\
=\left( 1-\left( \widetilde{\beta ^{i}}(g,x^{5})\right) ^{2}\right)
^{-1/2}=\left( 1-\left( \dfrac{v^{i}(g)b_{i}(x^{5})}{cb_{0}(x^{5})}\right)
^{2}\right) ^{-1/2} \\
\\
\\
II)\text{ \ }b_{0}^{-1}(x^{5})b_{i}(x^{5})\left( \sinh \tilde{\zeta}%
_{i}(g,x^{5})\right) =\widetilde{\gamma }(g,x^{5})\widetilde{\beta ^{i}}%
(g,x^{5})\dfrac{b_{i}(x^{5})}{b_{0}(x^{5})}= \\
\\
=\dfrac{v^{i}(g)b_{i}^{2}(x^{5})}{cb_{0}^{2}(x^{5})}\left( 1-\left( \dfrac{%
v^{i}(g)b_{i}(x^{5})}{cb_{0}(x^{5})}\right) ^{2}\right) ^{-1/2} \\
\\
\\
III)\text{ \ }b_{0}(x^{5})b_{i}^{-1}(x^{5})\left( \sinh \tilde{\zeta}%
_{i}(g,x^{5})\right) =\widetilde{\gamma }(g,x^{5})\widetilde{\beta ^{i}}%
(g,x^{5})\dfrac{b_{0}(x^{5})}{b_{i}(x^{5})}= \\
\\
=\dfrac{v^{i}(g)}{c}\left( 1-\left( \dfrac{v^{i}(g)b_{i}(x^{5})}{%
cb_{0}(x^{5})}\right) ^{2}\right) ^{-1/2}\bigskip .
\end{array}
\right. \medskip  \nonumber \\
\end{gather}

From the above system one gets (ESC off):
\begin{equation}
\left( 1-\left( \frac{v^{i}(g)b_{i}(x^{5})}{cb_{0}(x^{5})}\right)
^{2}\right) ^{-1/2}=\widetilde{\gamma }(g,x^{5})\equiv \widetilde{\gamma }%
^{i}(g,x^{5})=\cosh \tilde{\zeta}_{i}(g,x^{5});
\end{equation}
\begin{gather}
\frac{v^{i}(g)b_{i}(x^{5})}{cb_{0}(x^{5})}\left( 1-\left( \frac{%
v^{i}(g)b_{i}(x^{5})}{cb_{0}(x^{5})}\right) ^{2}\right) ^{-1/2}=\widetilde{%
\gamma }(g,x^{5})\widetilde{\beta ^{i}}(g,x^{5})\equiv \widetilde{\gamma }%
^{i}(g,x^{5})\widetilde{\beta ^{i}}(g,x^{5})=  \nonumber \\
\nonumber \\
=\sinh \tilde{\zeta}_{i}(g,x^{5}).
\end{gather}

Such relations are consistent with the properties of hyperbolic functions,
since ($\forall i=1,2,3$) (ESC off)
\begin{gather}
\cosh ^{2}\tilde{\zeta}_{i}(g,x^{5})-\sinh ^{2}\tilde{\zeta}%
_{i}(g,x^{5})=1\Leftrightarrow  \nonumber \\
\nonumber \\
\Leftrightarrow \left( 1-\left( \frac{v^{i}(g)b_{i}(x^{5})}{cb_{0}(x^{5})}%
\right) ^{2}\right) ^{-1}-\left( \frac{v^{i}(g)b_{i}(x^{5})}{cb_{0}(x^{5})}%
\right) ^{2}\left( 1-\left( \frac{v^{i}(g)b_{i}(x^{5})}{cb_{0}(x^{5})}%
\right) ^{2}\right) ^{-1}=  \nonumber \\
\nonumber \\
=\frac{\left( 1-\left( \dfrac{v^{i}(g)b_{i}(x^{5})}{cb_{0}(x^{5})}\right)
^{2}\right) }{\left( 1-\left( \dfrac{v^{i}(g)b_{i}(x^{5})}{cb_{0}(x^{5})}%
\right) ^{2}\right) }=1\smallskip .
\end{gather}
Eqs. (55) and (56) reduce of course to the standard SR relations in the
limit $\widetilde{M_{4}}(x^{5})$ $\longrightarrow $ $M_{4}$.

\subsection{Finite Deformed Rotation About a Coordinate Axis \ \ \ }

Let us now consider a finite true (clockwise) deformed rotation by an angle $%
\theta _{1}(g)$ about $\widehat{x^{1}}$. By recalling that ${\bf S}%
_{DSR4}(x^{5})\equiv
(I_{DSR4}^{23}(x^{5}),I_{DSR4}^{31}(x^{5}),I_{DSR4}^{12}(x^{5}))$, it
follows:\bigskip\
\begin{gather}
SO(3,1)_{DEF.}\ni g:x^{\mu }\rightarrow x_{(g)}^{\mu \prime }(\left\{
x\right\} _{m.},x^{5})=  \nonumber \\
\nonumber \\
=\exp (-\theta _{1}(g)S_{DSR4}^{1}(x^{5}))_{~~\nu }^{\mu }x^{\nu }=\exp
(\theta _{1}(g)I_{DSR4}^{32}(x^{5}))_{~~\nu }^{\mu }x^{\nu }=  \nonumber \\
\nonumber \\
=\left( \sum_{n=0}^{\infty }\frac{1}{n!}(\theta
_{1}(g)I_{DSR4}^{32}(x^{5}))^{n}\right) _{~~\nu }^{\mu }x^{\nu }=\left(
\sum_{n=0}^{\infty }\frac{1}{n!}(\theta
_{1}(g))^{n}(I_{DSR4}^{32}(x^{5}))^{n}\right) _{~~\nu }^{\mu }x^{\nu }=
\nonumber \\
\nonumber \\
=\left( 1_{4-d.}+\sum_{n=1}^{\infty }\frac{1}{\left( 2n\right) !}(\theta
_{1}(g))^{2n}(I_{DSR4}^{32}(x^{5}))^{2n}+\right.  \nonumber \\
\nonumber \\
\left. +\sum_{n=0}^{\infty }\frac{1}{\left( 2n+1\right) !}(\theta
_{1}(g))^{2n+1}(I_{DSR4}^{32}(x^{5}))_{~~}^{2n+1}\right) _{~~\nu }^{\mu
}x^{\nu }=  \nonumber \\
\nonumber \\
=\left( 1_{4-d.}+B_{1}\sum_{n=1}^{\infty }\frac{(-1)^{n}}{\left( 2n\right) !}%
b_{2}^{-2n}(x^{5})b_{3}^{-2n}(x^{5})(\theta _{1}(g))^{2n}+\right.  \nonumber
\\
\nonumber \\
\left. +I_{DSR4}^{32}(x^{5})\sum_{n=0}^{\infty }\frac{(-1)^{n}}{\left(
2n+1\right) !}b_{2}^{-2n}(x^{5})b_{3}^{-2n}(x^{5})(\theta
_{1}(g))^{2n+1})\right) _{~~\nu }^{\mu }x^{\nu }=  \nonumber \\
\nonumber \\
=\left( 1_{4-d.}+B_{1}\sum_{n=1}^{\infty }\frac{(-1)^{n}}{\left( 2n\right) !}%
(\theta _{1}(g)b_{2}^{-1}(x^{5})b_{3}^{-1}(x^{5}))^{2n}+\right.  \nonumber \\
\nonumber \\
\left. +I_{DSR4}^{32}(x^{5})b_{2}(x^{5})b_{3}(x^{5})\sum_{n=0}^{\infty }%
\frac{(-1)^{n}}{\left( 2n+1\right) !}(\theta
_{1}(g)b_{2}^{-1}(x^{5})b_{3}^{-1}(x^{5}))^{2n+1}\right) _{~~\nu }^{\mu
}x^{\nu }=  \nonumber \\
\nonumber \\
=\left( 1_{4-d.}+\left( \left( \cos \theta
_{1}(g)b_{2}^{-1}(x^{5})b_{3}^{-1}(x^{5})\right) -1\right) B_{1}+\right.
\nonumber \\
\nonumber \\
\left. +b_{2}(x^{5})b_{3}(x^{5})\left( \sin \theta
_{1}(g)b_{2}^{-1}(x^{5})b_{3}^{-1}(x^{5})\right) I_{DSR4}^{32}(x^{5})\right)
_{~~\nu }^{\mu }x^{\nu }\smallskip ,
\end{gather}
where in the last passage the series expansions of trigonometric functions
have been used.

By introducing the matrix
\begin{gather}
Rot._{DSR4,\widehat{x^{1}}}(g,x^{5})\equiv 1_{4-d.}+\left( \left( \cos
\theta _{1}(g)b_{2}^{-1}(x^{5})b_{3}^{-1}(x^{5})\right) -1\right) B_{1}+
\nonumber \\
\nonumber \\
+b_{2}(x^{5})b_{3}(x^{5})\left( \sin \theta
_{1}(g)b_{2}^{-1}(x^{5})b_{3}^{-1}(x^{5})\right) I_{DSR4}^{32}(x^{5}),
\end{gather}
Eq. (59) can be rewritten as:
\begin{gather}
SO(3,1)_{DEF.}\ni g:x^{\mu }\rightarrow \left( x^{\prime }\right)
_{(g),DSR4}^{\mu }(\left\{ x\right\} _{m.},x^{5})=\left( x^{\mu }\right)
_{(g),DSR4}^{^{\prime }}(\left\{ x\right\} _{m.},x^{5})=  \nonumber \\
\nonumber \\
=\left( Rot._{DSR4,\widehat{x^{1}}}(g,x^{5})\smallskip \right) _{~~\nu
}^{\mu }x^{\nu }.
\end{gather}
We finally have:
\begin{gather}
\left(
\begin{array}{c}
\left( x^{\prime }\right) _{(g),DSR4}^{0}(\left\{ x\right\} _{m.},x^{5}) \\
\left( x^{\prime }\right) _{(g),DSR4}^{1}(\left\{ x\right\} _{m.},x^{5}) \\
\left( x^{\prime }\right) _{(g),DSR4}^{2}(\left\{ x\right\} _{m.},x^{5}) \\
\left( x^{\prime }\right) _{(g),DSR4}^{3}(\left\{ x\right\} _{m.},x^{5})
\end{array}
\right) =\left(
\begin{array}{c}
\left( x^{0}\right) _{(g),DSR4}^{^{\prime }}(\left\{ x\right\} _{m.},x^{5})
\\
\left( x^{1}\right) _{(g),DSR4}^{^{\prime }}(\left\{ x\right\} _{m.},x^{5})
\\
\left( x^{2}\right) _{(g),DSR4}^{^{\prime }}(\left\{ x\right\} _{m.},x^{5})
\\
\left( x^{3}\right) _{(g),DSR4}^{^{\prime }}(\left\{ x\right\} _{m.},x^{5})
\end{array}
\right) =  \nonumber \\
\nonumber \\
=\left( Rot._{DSR4,\widehat{x^{1}}}(g,x^{5})\smallskip \right) \times \left(
\begin{array}{c}
x^{0} \\
x^{1} \\
x^{2} \\
x^{3}
\end{array}
\right) =  \nonumber \\
\nonumber \\
=\left(
\begin{array}{c}
x^{0} \\
\\
x^{1} \\
\\
(\cos \theta
_{1}(g)b_{2}^{-1}(x^{5})b_{3}^{-1}(x^{5}))x^{2}+b_{2}^{-1}(x^{5})b_{3}(x^{5})\left( \sin \theta _{1}(g)b_{2}^{-1}(x^{5})b_{3}^{-1}(x^{5})\right) x^{3}
\\
\\
-b_{2}(x^{5})b_{3}^{-1}(x^{5})\left( \sin \theta
_{1}(g)b_{2}^{-1}(x^{5})b_{3}^{-1}(x^{5})\right) x^{2}+(\cos \theta
_{1}(g)b_{2}^{-1}(x^{5})b_{3}^{-1}(x^{5}))x^{3}
\end{array}
\right) .\smallskip  \nonumber \\
\end{gather}
Analogous relations hold for the finite true (clockwise) deformed rotations
about the other two spatial axes. One has\
\begin{gather}
SO(3,1)_{DEF.}\ni g:x^{\mu }\rightarrow \left( x^{\prime }\right)
_{(g),DSR4}^{\mu }(\left\{ x\right\} _{m.},x^{5})=\left( x^{\mu }\right)
_{(g),DSR4}^{^{\prime }}(\left\{ x\right\} _{m.},x^{5})=  \nonumber \\
\nonumber \\
=\left( Rot._{DSR4,\widehat{x^{2}}}(g,x^{5})\smallskip \right) _{~~\nu
}^{\mu }x^{\nu };  \nonumber \\
\end{gather}
\begin{gather}
SO(3,1)_{DEF.}\ni g:x^{\mu }\rightarrow \left( x^{\prime }\right)
_{(g),DSR4}^{\mu }(\left\{ x\right\} _{m.},x^{5})=\left( x^{\mu }\right)
_{(g),DSR4}^{^{\prime }}(\left\{ x\right\} _{m.},x^{5})=  \nonumber \\
\nonumber \\
=\left( Rot._{DSR4,\widehat{x^{3}}}(g,x^{5})\smallskip \right) _{~~\nu
}^{\mu }x^{\nu },  \nonumber \\
\end{gather}
where we defined
\begin{gather}
Rot._{DSR4,\widehat{x^{2}}}(g,x^{5})\equiv 1_{4-d.}+\left( \left( \cos
\theta _{2}(g)b_{1}^{-1}(x^{5})b_{3}^{-1}(x^{5})\right) -1\right) B_{2}+
\nonumber \\
\nonumber \\
+b_{1}(x^{5})b_{3}(x^{5})\left( \sin \theta
_{2}(g)b_{1}^{-1}(x^{5})b_{3}^{-1}(x^{5})\right) I_{DSR4}^{13}(x^{5});
\nonumber \\
\end{gather}
\begin{gather}
Rot._{DSR4,\widehat{x^{3}}}(g,x^{5})\equiv 1_{4-d.}+\left( \left( \cos
\theta _{3}(g)b_{1}^{-1}(x^{5})b_{2}^{-1}(x^{5})\right) -1\right) B_{3}+
\nonumber \\
\nonumber \\
+b_{1}(x^{5})b_{2}(x^{5})\left( \sin \theta
_{3}(g)b_{1}^{-1}(x^{5})b_{2}^{-1}(x^{5})\right) I_{DSR4}^{21}(x^{5}).
\nonumber \\
\end{gather}
\bigskip

By introducing the {\em (effective) deformed angles }${\bf \tilde{\theta}}%
(g) $ ($i\neq j$, $i\neq k$, $j\neq k$)\footnote{%
Definition (67) of {\em \ }${\bf \tilde{\theta}}(g)$ does formally coincides
with that of the isotopic angles introduced by Santilli [11].
\par
{}}\bigskip\
\begin{gather}
\tilde{\theta}_{i}(g,x^{5})\equiv \theta
_{i}(g)b_{j}^{-1}(x^{5})b_{k}^{-1}(x^{5})=  \nonumber \\
\nonumber \\
\stackrel{\text{ESC off on }i\text{, ESC on on }j\text{\ and }k\text{%
{\footnotesize \ }}}{=}\frac{1}{2}\theta _{i}(g)\left| \epsilon
_{ijk}\right| b_{j}^{-1}(x^{5})b_{k}^{-1}(x^{5}),\text{ }\forall i=1,2,3%
\text{ ,}
\end{gather}
\bigskip Eqs. (60), (61) and (63)-(66) can be written in compact form
respectively as
\begin{gather}
\left(
\begin{array}{c}
\left( x^{\prime }\right) _{(g),DSR4}^{0}(\left\{ x\right\} _{m.},x^{5}) \\
\left( x^{\prime }\right) _{(g),DSR4}^{1}(\left\{ x\right\} _{m.},x^{5}) \\
\left( x^{\prime }\right) _{(g),DSR4}^{2}(\left\{ x\right\} _{m.},x^{5}) \\
\left( x^{\prime }\right) _{(g),DSR4}^{3}(\left\{ x\right\} _{m.},x^{5})
\end{array}
\right) =\left(
\begin{array}{c}
\left( x^{0}\right) _{(g),DSR4}^{^{\prime }}(\left\{ x\right\} _{m.},x^{5})
\\
\left( x^{1}\right) _{(g),DSR4}^{^{\prime }}(\left\{ x\right\} _{m.},x^{5})
\\
\left( x^{2}\right) _{(g),DSR4}^{^{\prime }}(\left\{ x\right\} _{m.},x^{5})
\\
\left( x^{3}\right) _{(g),DSR4}^{^{\prime }}(\left\{ x\right\} _{m.},x^{5})
\end{array}
\right) =  \nonumber \\
\nonumber \\
=\left(
\begin{array}{c}
x^{0}\medskip \\
\\
x^{1}\medskip \\
\\
(\cos \tilde{\theta}_{1}(g,x^{5}))x^{2}+b_{2}^{-1}(x^{5})b_{3}(x^{5})\left(
\sin \tilde{\theta}_{1}(g,x^{5})\right) x^{3}\medskip \\
\\
-b_{2}(x^{5})b_{3}^{-1}(x^{5})\left( \sin \tilde{\theta}_{1}(g,x^{5})\right)
x^{2}+(\cos \tilde{\theta}_{1}(g,x^{5}))x^{3}\medskip
\end{array}
\right) ;
\end{gather}
\begin{gather}
\left(
\begin{array}{c}
\left( x^{\prime }\right) _{(g),DSR4}^{0}(\left\{ x\right\} _{m.},x^{5}) \\
\left( x^{\prime }\right) _{(g),DSR4}^{1}(\left\{ x\right\} _{m.},x^{5}) \\
\left( x^{\prime }\right) _{(g),DSR4}^{2}(\left\{ x\right\} _{m.},x^{5}) \\
\left( x^{\prime }\right) _{(g),DSR4}^{3}(\left\{ x\right\} _{m.},x^{5})
\end{array}
\right) =\left(
\begin{array}{c}
\left( x^{0}\right) _{(g),DSR4}^{^{\prime }}(\left\{ x\right\} _{m.},x^{5})
\\
\left( x^{1}\right) _{(g),DSR4}^{^{\prime }}(\left\{ x\right\} _{m.},x^{5})
\\
\left( x^{2}\right) _{(g),DSR4}^{^{\prime }}(\left\{ x\right\} _{m.},x^{5})
\\
\left( x^{3}\right) _{(g),DSR4}^{^{\prime }}(\left\{ x\right\} _{m.},x^{5})
\end{array}
\right) =  \nonumber \\
\nonumber \\
=\left(
\begin{array}{c}
x^{0}\medskip \\
\\
x^{1}\medskip \\
\\
(\cos \tilde{\theta}_{1}(g,x^{5}))x^{2}+b_{2}^{-1}(x^{5})b_{3}(x^{5})\left(
\sin \tilde{\theta}_{1}(g,x^{5})\right) x^{3}\medskip \\
\\
-b_{2}(x^{5})b_{3}^{-1}(x^{5})\left( \sin \tilde{\theta}_{1}(g,x^{5})\right)
x^{2}+(\cos \tilde{\theta}_{1}(g,x^{5}))x^{3}\medskip
\end{array}
\right) ;
\end{gather}
\begin{gather}
\left(
\begin{array}{c}
\left( x^{\prime }\right) _{(g),DSR4}^{0}(\left\{ x\right\} _{m.},x^{5}) \\
\left( x^{\prime }\right) _{(g),DSR4}^{1}(\left\{ x\right\} _{m.},x^{5}) \\
\left( x^{\prime }\right) _{(g),DSR4}^{2}(\left\{ x\right\} _{m.},x^{5}) \\
\left( x^{\prime }\right) _{(g),DSR4}^{3}(\left\{ x\right\} _{m.},x^{5})
\end{array}
\right) =\left(
\begin{array}{c}
\left( x^{0}\right) _{(g),DSR4}^{^{\prime }}(\left\{ x\right\} _{m.},x^{5})
\\
\left( x^{1}\right) _{(g),DSR4}^{^{\prime }}(\left\{ x\right\} _{m.},x^{5})
\\
\left( x^{2}\right) _{(g),DSR4}^{^{\prime }}(\left\{ x\right\} _{m.},x^{5})
\\
\left( x^{3}\right) _{(g),DSR4}^{^{\prime }}(\left\{ x\right\} _{m.},x^{5})
\end{array}
\right) =  \nonumber \\
\nonumber \\
=\left(
\begin{array}{c}
x^{0}\medskip \\
\\
(\cos \tilde{\theta}_{3}(g,x^{5}))x^{1}+b_{1}^{-1}(x^{5})b_{2}(x^{5})\left(
\sin \tilde{\theta}_{3}(g,x^{5})\right) x^{2}\medskip \\
\\
-b_{1}(x^{5})b_{2}^{-1}(x^{5})\left( \sin \tilde{\theta}_{3}(g,x^{5})\right)
x^{1}+(\cos \tilde{\theta}_{3}(g,x^{5}))x^{2}\medskip \\
\\
x^{3}\medskip
\end{array}
\right) .
\end{gather}

By comparing the finite true (clockwise) rotations by an angle $\theta
^{i}(g)$ about $\widehat{x^{i}}$ in SR
\begin{gather}
\left(
\begin{array}{c}
\left( x^{\prime }\right) _{(g),SR4}^{0}(\left\{ x\right\} _{m.}) \\
\left( x^{\prime }\right) _{(g),SR4}^{1}(\left\{ x\right\} _{m.}) \\
\left( x^{\prime }\right) _{(g),SR4}^{2}(\left\{ x\right\} _{m.}) \\
\left( x^{\prime }\right) _{(g),SR4}^{3}(\left\{ x\right\} _{m.})
\end{array}
\right) =\left(
\begin{array}{c}
\left( x^{0}\right) _{(g),SR4}^{^{\prime }}(\left\{ x\right\} _{m.}) \\
\left( x^{1}\right) _{(g),SR4}^{^{\prime }}(\left\{ x\right\} _{m.}) \\
\left( x^{2}\right) _{(g),SR4}^{^{\prime }}(\left\{ x\right\} _{m.}) \\
\left( x^{3}\right) _{(g),SR4}^{^{\prime }}(\left\{ x\right\} _{m.})
\end{array}
\right) =  \nonumber \\
\nonumber \\
=\left(
\begin{array}{c}
x^{0}\smallskip \\
x^{1}\smallskip \\
\left( \cos \theta _{1}(g)\right) x^{2}+\left( \sin \theta _{1}(g)\right)
x^{3}\smallskip \\
-\left( \sin \theta _{1}(g)\right) x^{2}+\left( \cos \theta _{1}(g)\right)
x^{3}\smallskip
\end{array}
\right) ;  \nonumber \\
\end{gather}
\begin{gather}
\left(
\begin{array}{c}
\left( x^{\prime }\right) _{(g),SR4}^{0}(\left\{ x\right\} _{m.}) \\
\left( x^{\prime }\right) _{(g),SR4}^{1}(\left\{ x\right\} _{m.}) \\
\left( x^{\prime }\right) _{(g),SR4}^{2}(\left\{ x\right\} _{m.}) \\
\left( x^{\prime }\right) _{(g),SR4}^{3}(\left\{ x\right\} _{m.})
\end{array}
\right) =\left(
\begin{array}{c}
\left( x^{0}\right) _{(g),SR4}^{^{\prime }}(\left\{ x\right\} _{m.}) \\
\left( x^{1}\right) _{(g),SR4}^{^{\prime }}(\left\{ x\right\} _{m.}) \\
\left( x^{2}\right) _{(g),SR4}^{^{\prime }}(\left\{ x\right\} _{m.}) \\
\left( x^{3}\right) _{(g),SR4}^{^{\prime }}(\left\{ x\right\} _{m.})
\end{array}
\right) =  \nonumber \\
\nonumber \\
=\left(
\begin{array}{c}
x^{0}\smallskip \\
\left( \cos \theta _{2}(g)\right) x^{1}-\left( \sin \theta _{2}(g)\right)
x^{3}\smallskip \\
x^{2}\smallskip \\
\left( \sin \theta _{2}(g)\right) x^{1}+\left( \cos \theta _{2}(g)\right)
x^{3}\smallskip
\end{array}
\right) ;  \nonumber \\
\end{gather}
\begin{gather}
\left(
\begin{array}{c}
\left( x^{\prime }\right) _{(g),SR4}^{0}(\left\{ x\right\} _{m.}) \\
\left( x^{\prime }\right) _{(g),SR4}^{1}(\left\{ x\right\} _{m.}) \\
\left( x^{\prime }\right) _{(g),SR4}^{2}(\left\{ x\right\} _{m.}) \\
\left( x^{\prime }\right) _{(g),SR4}^{3}(\left\{ x\right\} _{m.})
\end{array}
\right) =\left(
\begin{array}{c}
\left( x^{0}\right) _{(g),SR4}^{^{\prime }}(\left\{ x\right\} _{m.}) \\
\left( x^{1}\right) _{(g),SR4}^{^{\prime }}(\left\{ x\right\} _{m.}) \\
\left( x^{2}\right) _{(g),SR4}^{^{\prime }}(\left\{ x\right\} _{m.}) \\
\left( x^{3}\right) _{(g),SR4}^{^{\prime }}(\left\{ x\right\} _{m.})
\end{array}
\right) =  \nonumber \\
\nonumber \\
=\left(
\begin{array}{c}
x^{0}\smallskip \\
\left( \cos \theta _{3}(g)\right) x^{1}+\left( \sin \theta _{3}(g)\right)
x^{2}\smallskip \\
-\left( \sin \theta _{3}(g)\right) x^{1}+\left( \cos \theta _{3}(g)\right)
x^{2}\smallskip \\
x^{3}\smallskip
\end{array}
\right)  \nonumber \\
\end{gather}
with those in DSR4, it is easily seen that the deforming transition SR$%
\rightarrow $DSR4 corresponds --- at the level of group parameters --- to
the {\em deforming and anisotropizing rescaling} of angles $\theta
_{i}(g)\rightarrow \widetilde{\theta }_{i}(g,x^{5})$, $\forall i=1,2,3$.

Let us notice that the chronotopical rotation group $SO(3,1)_{GEN.}$ of the
generalized 4-d. ($(3,1)$ hyperbolically-signed) Minkowski space is
non-compact (like the standard one $SO(3,1)$ of usual SR). This is obviously
related to the existence of at least one timelike dimension, namely of
pseudo-rotations (or boost transformations). Indeed, \ whereas the range of
the 3-vector angle parameter ${\bf \theta }(g)$ is compact ($\theta
^{i}(g)\in \lbrack 0,2\pi ]$ \ $\forall i=1,2,3$), each component of the
rapidity 3-vector ${\bf \zeta }(g)$ has a non-compact range (the whole real
line: $\zeta ^{i}(g)\in R\equiv (-\infty ,+\infty )$ \ $\forall i=1,2,3$)%
\footnote{%
The same holds true of course for the velocity parameter. In SR, the light
speed $c$ being the maximal causal velocity (m.c.v.) implies that the range
of the dimensional boost parameter $v^{i}(g)$ is the real, non-compact
(since bounded but open) interval $(-c,+c)$ $\forall g\in
SO(3,1)_{STD.}\medskip $ (namely, ''luminal'' boosts are not allowed, what
amounts to say that no rest frame exists for a massless particle.).
Analogously, in DSR4, where the m.c.v. is given by $u_{DSR4}^{i}=\frac{%
b_{0}(x^{5})}{b_{i}(x^{5})}c$, the range of the dimensional velocity
parameter $v^{i}(g)$ of the deformed boost is the real, non-compact interval
$\left( -u_{DSR4}^{i}=-c\frac{b_{0}(x^{5})}{b_{i}(x^{5})},+u_{DSR4}^{i}=+c%
\frac{b_{0}(x^{5})}{b_{i}(x^{5})}\right) $:
\[
v^{i}(g)\in (-u_{DSR4}^{i},+u_{DSR4}^{i}),\text{ }\forall i=1,2,3,\text{ }%
\forall g\in SO(3,1)_{DEF.}\medskip ,
\]
since
\[
\widetilde{\beta ^{i}}(g)|_{v^{i}(g)=\pm u_{DSR4}^{i}}=\pm \frac{u_{DSR4}^{i}%
}{u_{DSR4}^{i}}=\pm 1;
\]
\par
\begin{eqnarray*}
\widetilde{\gamma ^{i}}(g)|_{v^{i}(g)=\pm u_{DSR4}^{i}} &=&\left[ \left(
1-\left( \widetilde{\beta ^{i}}(g)\right) ^{2}\right) ^{-1/2}\right]
|_{v^{i}(g)=\pm u_{DSR4}^{i}}= \\
&& \\
&=&1-\left[ \left( \widetilde{\beta ^{i}}(g)|_{v^{i}(g)=\pm
u_{DSR4}^{i}}\right) ^{2}\right] ^{-1/2}=\infty \medskip .
\end{eqnarray*}
\par
{}}.

Such a conclusion holds true, in general, for the $N$-dimensional case. The
presence of timelike dimensions (i.e. the fact $T>0$), and therefore of true
space-time mixing, implies the lack of compactness of the chronotopical
rotation group of the $N$-d. generalized Minkowski space, i.e. of the
homogeneous component of the corresponding (maximal) Killing group $%
P(S,T)_{GEN.}$.

\subsection{The Antisymmetric Tensor $\protect\delta \protect\omega _{DSR4}^{%
\protect\mu \protect\nu }(g,x^{5})$ of the Dimensionless (Effective)
Deformed Parameters of the Space-Time Rotation Group in DSR4}

It was shown in Ref. [1] that, for a generalized Minkowski space, {\em all}
forms of the tensor $\delta \omega $ are {\em global}, i.e. independent of
the set of metric variables $\left\{ x\right\} _{m.}\stackrel{S=3,\text{ }T=1%
\text{ case}}{=}\left\{ x^{0},x^{1},x^{2},x^{3}\right\} $, but only \ $%
\delta \omega _{\alpha \beta }(g)$ is {\em a priori} independent of possible
{\em non-metric} variables $\left\{ x\right\} _{n.m.}$. For instance,
consider the completely contravariant form of $\delta \omega $:
\begin{equation}
\delta \omega ^{\alpha \beta }(g,\left\{ x\right\} _{n.m.})\equiv g^{\alpha
\gamma }(\left\{ x\right\} _{n.m.})g^{\beta \delta }(\left\{ x\right\}
_{n.m.})\delta \omega _{\alpha \beta }(g)
\end{equation}
or, in matrix form:

\begin{equation}
\delta \omega _{contrav.}(g,\left\{ x\right\} _{n.m.})\equiv
g_{contrav.}^{T}(\left\{ x\right\} _{n.m.})\times \delta \omega
_{cov.}(g)\times g_{contrav.}(\left\{ x\right\} _{n.m.}).
\end{equation}

In the DSR4 case, the completely contravariant metric tensor reads
\begin{gather}
g_{DSR4}^{\mu \nu
}(x^{5})=diag(b_{0}^{-2}(x^{5}),-b_{1}^{-2}(x^{5}),-b_{2}^{-2}(x^{5}),-b_{3}^{-2}(x^{5}))=
\nonumber \\
\nonumber \\
\stackrel{ESCoff}{=}\delta ^{\mu \nu }(\delta ^{\mu
0}b_{0}^{-2}(x^{5})-\delta ^{\mu 1}b_{1}^{-2}(x^{5})-\delta ^{\mu
2}b_{2}^{-2}(x^{5})-\delta ^{\mu 3}b_{3}^{-2}(x^{5}))\smallskip
\end{gather}
or
\begin{equation}
g_{contrav.DSR4}(x^{5})=\left(
\begin{array}{cccc}
b_{0}^{-2}(x^{5}) & 0 & 0 & 0 \\
0 & -b_{1}^{-2}(x^{5}) & 0 & 0 \\
0 & 0 & -b_{2}^{-2}(x^{5}) & 0 \\
0 & 0 & 0 & -b_{3}^{-2}(x^{5})
\end{array}
\right) .
\end{equation}
Therefore
\begin{gather}
\delta \omega _{DSR4}^{\alpha \beta }(g,x^{5})\equiv \delta \omega
_{contrav.,DSR4}(g,x^{5})\equiv  \nonumber \\
\nonumber \\
\nonumber \\
\equiv g_{contrav.DSR4}^{T}(x^{5})\times \delta \omega
_{cov.,(DSR4)}(g)\times g_{contrav.DSR4}(x^{5})=  \nonumber \\
\nonumber \\
\nonumber \\
=\left(
\begin{array}{cccc}
b_{0}^{-2}(x^{5}) & 0 & 0 & 0 \\
0 & -b_{1}^{-2}(x^{5}) & 0 & 0 \\
0 & 0 & -b_{2}^{-2}(x^{5}) & 0 \\
0 & 0 & 0 & -b_{3}^{-2}(x^{5})
\end{array}
\right) \times  \nonumber \\
\nonumber \\
\times \left(
\begin{array}{cccc}
0 & -\zeta ^{1}(g) & -\zeta ^{2}(g) & -\zeta ^{3}(g) \\
\zeta ^{1}(g) & 0 & -\theta ^{3}(g) & \theta ^{2}(g) \\
\zeta ^{2}(g) & \theta ^{3}(g) & 0 & -\theta ^{1}(g) \\
\zeta ^{3}(g) & -\theta ^{2}(g) & \theta ^{1}(g) & 0
\end{array}
\right) \times  \nonumber \\
\nonumber \\
\times \left(
\begin{array}{cccc}
b_{0}^{-2}(x^{5}) & 0 & 0 & 0 \\
0 & -b_{1}^{-2}(x^{5}) & 0 & 0 \\
0 & 0 & -b_{2}^{-2}(x^{5}) & 0 \\
0 & 0 & 0 & -b_{3}^{-2}(x^{5})
\end{array}
\right) =  \nonumber \\
\nonumber \\
\nonumber \\
=\left(
\begin{array}{cccc}
b_{0}^{-2}(x^{5}) & 0 & 0 & 0 \\
0 & -b_{1}^{-2}(x^{5}) & 0 & 0 \\
0 & 0 & -b_{2}^{-2}(x^{5}) & 0 \\
0 & 0 & 0 & -b_{3}^{-2}(x^{5})
\end{array}
\right) \times  \nonumber \\
\\
\times \left(
\begin{array}{cccc}
0 & \zeta ^{1}(g)b_{1}^{-2}(x^{5}) & \zeta ^{2}(g)b_{2}^{-2}(x^{5}) & \zeta
^{3}(g)b_{3}^{-2}(x^{5}) \\
\zeta ^{1}(g)b_{0}^{-2}(x^{5}) & 0 & \theta ^{3}(g)b_{2}^{-2}(x^{5}) &
-\theta ^{2}(g)b_{3}^{-2}(x^{5}) \\
\zeta ^{2}(g)b_{0}^{-2}(x^{5}) & -\theta ^{3}(g)b_{1}^{-2}(x^{5}) & 0 &
\theta ^{1}(g)b_{3}^{-2}(x^{5}) \\
\zeta ^{3}(g)b_{0}^{-2}(x^{5}) & \theta ^{2}(g)b_{1}^{-2}(x^{5}) & -\theta
^{1}(g)b_{2}^{-2}(x^{5}) & 0
\end{array}
\right) ,  \nonumber
\end{gather}
whence, by recalling the definitions (44) and (67) of the (effective)
deformed rapidity and angle (Euclidean) 3-vectors, we obtain :
\begin{eqnarray}
&&\left\{
\begin{array}{l}
\delta \omega _{DSR4}^{0i}(g,x^{5})=\zeta
^{i}(g)b_{0}^{-2}(x^{5})b_{i}^{-2}(x^{5})=\widetilde{\zeta }%
^{i}(g,x^{5})b_{0}^{-1}(x^{5})b_{i}^{-1}(x^{5}),\text{ }\forall \text{\ }%
i=1,2,3\medskip ; \\
\\
\\
\\
\left.
\begin{array}{c}
\delta \omega _{DSR4}^{12}(g,x^{5})=-\theta
^{3}(g)b_{1}^{-2}(x^{5})b_{2}^{-2}(x^{5})=-\widetilde{\theta }%
^{3}(g,x^{5})\medskip b_{1}^{-1}(x^{5})b_{2}^{-1}(x^{5}) \\
\\
\delta \omega _{DSR4}^{13}(g,x^{5})=\theta
^{2}(g)b_{1}^{-2}(x^{5})b_{3}^{-2}(x^{5})\medskip =\widetilde{\theta }%
^{2}(g,x^{5})\medskip b_{1}^{-1}(x^{5})b_{3}^{-1}(x^{5})\medskip \\
\\
\delta \omega _{DSR4}^{23}(g,x^{5})=-\theta
^{1}(g)b_{2}^{-2}(x^{5})b_{3}^{-2}(x^{5})\medskip =-\widetilde{\theta }%
^{1}(g,x^{5})\medskip b_{2}^{-1}(x^{5})b_{3}^{-1}(x^{5})\medskip
\end{array}
\right\} \Rightarrow \\
\text{ } \\
\Rightarrow \delta \omega _{DSR4}^{jk}(g,x^{5})\stackrel{\text{ESC on on }i%
\text{, ESC off on }j\text{\ and }k\text{{\footnotesize \ }}}{=}\epsilon
_{ikj}\widetilde{\theta }^{i}(g,x^{5})b_{j}^{-1}(x^{5})b_{k}^{-1}(x^{5}), \\
\text{ } \\
\forall \left( j,k\right) \in \left( 1,2,3\right) .
\end{array}
\right.  \nonumber \\
&&
\end{eqnarray}

The above relation between the components of $\delta \omega _{\alpha \beta
,\left( DSR4\right) }(g)$ and those of $\delta \omega _{DSR4}^{\alpha \beta
}(g,x^{5})$ can be therefore written as $(i,j=1,2,3)$ (ESC off):
\begin{gather}
\delta \omega _{ij,\left( DSR4\right)
}(g)=b_{i}^{2}(x^{5})b_{j}^{2}(x^{5})\delta \omega _{DSR4}^{ij}(g,x^{5});
\nonumber \\
\nonumber \\
\delta \omega _{0i,\left( DSR4\right)
}(g)=-b_{0}^{-2}(x^{5})b_{i}^{-2}(x^{5})\delta \omega _{DSR4}^{0i}(g,x^{5}).
\end{gather}

We can therefore state that the formal {\em ''anisotropizing deforming''
transition} SR$\rightarrow $DSR4 is summarized, at the group parameter
level, by the passage from the antisymmetric, parametric {\em covariant }%
tensor $\delta \omega _{\alpha \beta ,\left( SSR4\right) }(g)$ \footnote{%
Notice that formally
\begin{equation}
\delta \omega _{\alpha \beta ,\left( SSR4\right) }(g)=\delta \omega _{\alpha
\beta ,\left( DSR4\right) }(g),  \tag{$\circ $}
\end{equation}
but the $g$'s belong to {\em different} space-time rotation groups. In the
l.h.s of $\left( \circ \right) $ $g\in SO(3,1)_{STD.}$ (homogeneous Lorentz
group), while in the r.h.s. of $\left( \circ \right) $ $g$ belong to the
''deformed'' counterpart, i.e. $g\in SO(3,1)_{DEF.}$ (homogeneous
''deformed'' Lorentz group).} to the antisymmetric {\em contravariant}
tensor $\delta \omega _{DSR4}^{\alpha \beta }(g,x^{5})$ of the (effective)
deformed parameters of $SO(3,1)_{DEF.}$ \footnote{%
Let us notice that the same conclusion (namely, the characterization of $%
\delta \omega _{DSR4}^{\alpha \beta }(g,x^{5})$ as tensor of the effective
deformed parameters) does {\em not} hold at the infinitesimal level. Indeed,
in DSR4 (see Ref. [1]) --- and in SR as well --- one {\em cannot} rescale
parameters in a unique way, since in any fixed infinitesimal transformation
tipology the related group parameter is rescaled in a different way,
depending on the coordinate concerned. Effective deformed group parameters
can be therefore defined only at the finite level of the space-time rotation
component $SO(3,1)_{DEF.}$ of the maximal Killing group $P(3,1)_{DEF.}$ of
the deformed 4-d. Minkowski space $\widetilde{M_{4}}(x^{5})$.
\par
The possibility of generalizing such results to the generalized 4-d. $%
(S,T=4-S)$ (and possibly to the $N$-d. $(S,T=N-S)$) Minkowskian spaces
depends strongly on the explicit form of the metric tensor. For instance,
for a {\em non-diagonal} $N$-d. metric tensor, the parametric tensor $\delta
\omega ^{\alpha \beta }(g,\left\{ x\right\} _{n.m.})$\ in general {\em has
no definite symmetry property}, and therefore it is more difficult (or even
impossible) to identify its components with possible (effective) generalized
parameters.}.\pagebreak

\section{FINITE 3-d. DEFORMED TRUE ROTATIONS ABOUT\ A\ GENERIC\ AXIS IN DSR4}

\subsection{\protect\bigskip Parametric Decomposition}

We want now to derive the finite, deformed true (clockwise) rotations by a
generic angle $\varphi (g)$ about a generic axis $\widehat{\varepsilon }(g)$
in the physical 3-d. space $\widetilde{E_{3}}(x^{5})$ embedded in $%
\widetilde{M_{4}}(x^{5})$, by exploiting the form of the infinitesimal
generators of the DSR4 chronotopical group $SO(3,1)_{DEF.}$ obtained in Ref.
[1]. The metric structure in $\widetilde{E_{3}}(x^{5})$ is determined by $%
-g_{ij,DSR4}(x^{5})\stackrel{\text{{\footnotesize ESC off}}}{=}\delta
_{ij}b_{i}^{2}(x^{5})$ \footnote{%
The minus sign is introduced to obtain positive 3-vector norms $\forall
x^{5}\in R_{0}^{+}$.}. The unit vector of the rotation axis is
\[
\widehat{\varepsilon }(g)=\varepsilon ^{1}(g)\widehat{x^{1}}+\varepsilon
^{2}(g)\widehat{x^{2}}+\varepsilon ^{3}(g)\widehat{x^{3}},
\]
with
\[
|\widehat{\varepsilon }(g)|_{\ast }^{2}\equiv
\sum_{i=1}^{3}b_{i}^{2}(x^{5})\left( \varepsilon ^{i}(g)\right) ^{2}=1
\]
(the $\left| \cdot \right| _{\ast }$ is the 3-d. norm associated to $%
-g_{ij,DSR4}(x^{5})$).

In general, by the very basic properties of a group, it is {\em always}
possible to find a {\em (not unique)} axial-parametric decomposition which\
transfers--- once fixed the three coordinate axes in the space considered---
all the dependence on the group element $g$ only on the transformation
parameters $\left\{ \theta ^{i}(g)\right\} _{i=1,2,3}$
\begin{equation}
\left( \widehat{\varepsilon }(g),\varphi (g)\right) \rightarrow (\widehat{%
x^{1}},\theta ^{1}(g))(\widehat{x^{2}},\theta ^{2}(g))(\widehat{x^{3}}%
,\theta ^{3}(g)).
\end{equation}
At the infinitesimal (i.e. algebraic) level, such a decomposition is {\em %
independent} on the order, due to the commutativity of the infinitesimal
elements (i.e. of the transformations at an algebraic level) of any Lie
group of transformations. This of course does not longer hold at a finite
(i.e. group) level --- due to the non-Abelian nature of Lie groups of
chronotopical transformations---and therefore the order on the rhs of Eq.
(81) is fixed for a given pair $\left( \widehat{\varepsilon }(g),\varphi
(g)\right) $ \footnote{%
Needless to say, at the algebraic and group level rotation angles $\varphi
(g)$ and $\left\{ \theta ^{i}(g)\right\} _{i=1,2,3}$ are infinitesimal and
finite, respectively. This will be understood, and, for simplicity's sake,
no notational distinction will be made.}.

We have therefore, in DSR4 ($\cdot $ is the Euclidean scalar product):
\begin{gather}
-\varphi (g)\widehat{\varepsilon }(g)\cdot {\bf S}_{DSR4}(x^{5})\rightarrow
\nonumber \\
\nonumber \\
\rightarrow -\theta ^{1}(g)S_{DSR4}^{1}(x^{5})-\theta
^{2}(g)S_{DSR4}^{2}(x^{5})-\theta ^{3}(g)S_{DSR4}^{3}(x^{5})=  \nonumber \\
\medskip  \nonumber \\
=-\sum_{i=1}^{3}\theta ^{i}(g)S_{DSR4}^{i}(x^{5})\medskip
\end{gather}
(infinitesimal level, any composition order)\footnote{%
To be more precise, in Eq. (82) ---and in general for infinitesimal,
deformed true rotation transformations--- $g\in SO(3)_{DEF.}$ should be
replaced by $\delta g$, where $\delta g\in su(2)_{DEF.}$, i.e. it is an
element of the deformed true rotation algebra. But, for simplicity's sake,
we will omit, but mean, this cumbersome notation.};

\begin{gather}
\exp \left( -\varphi (g)\widehat{\varepsilon }(g)\cdot {\bf S}%
_{DSR4}(x^{5})\right) \rightarrow  \nonumber \\
\nonumber \\
\rightarrow \exp \left( -\theta ^{1}(g)S_{DSR4}^{1}(x^{5})\right) \times
\exp \left( -\theta ^{2}(g)S_{DSR4}^{2}(x^{5})\right) \times \exp \left(
-\theta ^{3}(g)S_{DSR4}^{3}(x^{5})\right)  \nonumber \\
\end{gather}
(finite level, fixed composition order).

A different axial-parametric decomposition utilizes the Euler
angles\linebreak\ $\left\{ \theta _{1}^{\overline{i}}(g)\text{, }\theta ^{^{%
\overline{j}\neq \overline{i}}}(g)\text{, }\theta _{2}^{^{\overline{i}%
}}(g)\right\} $:
\begin{gather}
\left( \widehat{\varepsilon }(g),\varphi (g)\right) \rightarrow (\widehat{x^{%
\overline{i}}},\theta _{1}^{\overline{i}}(g))(\widehat{x^{\overline{j}}}%
,\theta ^{^{\overline{j}}}(g))(\widehat{x^{^{\overline{i}}}},\theta _{2}^{^{%
\overline{i}}}(g));  \nonumber \\
\nonumber \\
\overline{i},\overline{j}\in \left\{ 1,2,3\right\} ,\overline{i}\neq
\overline{j}
\end{gather}
In such a case, only two coordinate space axes, $\widehat{x^{\overline{i}}}$
and $\widehat{x^{\overline{j}\neq \overline{i}}}$, are used, with a special
composition order. At infinitesimal level, one has
\begin{gather}
\left( \widehat{\varepsilon }(g),\varphi (g)\right) \rightarrow  \nonumber \\
\medskip  \nonumber \\
\rightarrow (\widehat{x^{\overline{i}}},\theta _{1}^{\overline{i}}(g))(%
\widehat{x^{\overline{j}\neq \overline{i}}},\theta ^{^{\overline{j}\neq
\overline{i}}}(g))(\widehat{x^{^{\overline{i}}}},\theta _{2}^{^{\overline{i}%
}}(g))\equiv (\widehat{x^{\overline{i}}},\theta _{1}^{\overline{i}%
}(g)+\theta _{2}^{^{\overline{i}}}(g))(\widehat{x^{^{\overline{i}}}},\theta
_{2}^{^{\overline{i}}}(g)).
\end{gather}
For DSR4, we have
\begin{gather}
-\varphi (g)\widehat{\varepsilon }(g)\cdot {\bf S}_{DSR4}(x^{5})\rightarrow
-\theta _{1}^{\overline{i}}(g)S_{DSR4}^{\overline{i}}(x^{5})-\theta ^{%
\overline{j}\neq \overline{i}}(g)S_{DSR4}^{\overline{j}\neq \overline{i}%
}(x^{5})-\theta _{2}^{\overline{i}}(g)S_{DSR4}^{\overline{i}}(x^{5})=
\nonumber \\
\medskip  \nonumber \\
=-\left( \theta _{1}^{\overline{i}}(g)+\theta _{2}^{\overline{i}}(g)\right)
S_{DSR4}^{\overline{i}}(x^{5})-\theta ^{\overline{j}\neq \overline{i}%
}(g)S_{DSR4}^{\overline{j}\neq \overline{i}}(x^{5})
\end{gather}
(infinitesimal level, any composition order);

\begin{gather}
\exp \left( -\varphi (g)\widehat{\varepsilon }(g)\cdot {\bf S}%
_{DSR4}(x^{5})\right) \rightarrow  \nonumber \\
\nonumber \\
\rightarrow \exp \left( -\theta _{1}^{\overline{i}}(g)S_{DSR4}^{\overline{i}%
}(x^{5})\right) \times \exp \left( -\theta ^{\overline{j}\neq \overline{i}%
}(g)S_{DSR4}^{\overline{j}\neq \overline{i}}(x^{5})\right) \times \exp
\left( -\theta _{2}^{\overline{i}}(g)S_{DSR4}^{\overline{i}}(x^{5})\right)
\medskip  \nonumber \\
\end{gather}
(finite level, fixed composition order).

It is also possible to find out by direct computation (namely, by
integration on the group parameters) the 4$\times $4 matrix representative
of the finite element $g$ of the rotation group $SO(3,1)_{DEF.}$ in DSR4
corresponding to a rotation by $\varphi (g)$ about $\widehat{\varepsilon }%
(g) $. We have
\begin{gather}
-\varphi (g)\widehat{\varepsilon }(g)\cdot {\bf S}_{DSR4}(x^{5})=-\varphi
(g)\sum_{i=1}^{3}\varepsilon ^{i}(g)S_{DSR4}^{i}(x^{5})\rightarrow  \nonumber
\\
\nonumber \\
\rightarrow \exp \left( -\varphi (g)\widehat{\varepsilon }(g)\cdot {\bf S}%
_{DSR4}(x^{5})\right) =\exp \left( -\varphi (g)\sum_{i=1}^{3}\varepsilon
^{i}(g)S_{DSR4}^{i}(x^{5})\right) \neq  \nonumber \\
\medskip  \nonumber \\
\neq \prod_{i=1}^{3}\exp \left( -\varphi (g)\varepsilon
^{i}(g)S_{DSR4}^{i}(x^{5})\right) \medskip ,
\end{gather}
where in the last passage the {\em Baker-Campbell-Hausdorff formula} [12] is
exploited,{\bf \ }the non-Abelian nature of $SO(3)_{DEF.}\subset
SO(3,1)_{DEF.}$ is used, and matrix product is understood in $%
\prod_{i=1}^{3} $.

Needless to say, all the above procedures are {\em equivalent} (although
they may yield different formal results). Let us exploit the first one.

\subsection{Exponentiating the Deformed Infinitesimal Rotation}

Let us denote by $A_{\left( \varphi (g),\widehat{\varepsilon }(g)\right)
,DSR4}(x^{5})$ the 4$\times $4 matrix corresponding to an infinitesimal
(clockwise) rotation by an (infinitesimal) angle $\varphi (g)$ about the
axis $\widehat{\varepsilon }(g)$ (matrix belonging a 4-d. representation of
the algebra of $SO(3)_{DEF.}$, namely $su(2)_{DEF.}$):
\begin{gather}
A_{\left( \varphi (g),\widehat{\varepsilon }(g)\right) ,DSR4}(x^{5})\equiv
-\varphi (g)\widehat{\varepsilon }(g)\cdot {\bf S}_{DSR4}(x^{5})\rightarrow
\nonumber \\
\nonumber \\
\rightarrow \left( -\theta ^{1}(g)S_{DSR4}^{1}(x^{5})-\theta
^{2}(g)S_{DSR4}^{2}(x^{5})-\theta ^{3}(g)S_{DSR4}^{3}(x^{5})\right)
\rightarrow  \nonumber \\
\medskip  \nonumber \\
\rightarrow \exp (A_{\left( \varphi (g),\widehat{\varepsilon }(g)\right)
,DSR4}(x^{5}))=  \nonumber \\
\medskip  \nonumber \\
=\exp \left( -\theta ^{1}(g)S_{DSR4}^{1}(x^{5})-\theta
^{2}(g)S_{DSR4}^{2}(x^{5})-\theta ^{3}(g)S_{DSR4}^{3}(x^{5})\right) =
\nonumber \\
\medskip  \nonumber \\
=\sum_{n=0}^{\infty }\frac{1}{n!}\left( -\theta
^{1}(g)S_{DSR4}^{1}(x^{5})-\theta ^{2}(g)S_{DSR4}^{2}(x^{5})-\theta
^{3}(g)S_{DSR4}^{3}(x^{5})\right) ^{n}\medskip ,
\end{gather}
where $\exp (A_{\left( \varphi (g),\widehat{\varepsilon }(g)\right)
,DSR4}(x^{5}))$ is the 4$\times $4 matrix corresponding to a finite
(clockwise) rotation by a (finite) angle $\varphi (g)$ about the axis $%
\widehat{\varepsilon }(g)$, belonging to a 4-d. representation of $%
SO(3)_{DEF.}$, of course.

On account of the explicit form of ${\bf S}_{DSR4}(x^{5})$ and of the
deformed rotation generators (see Subsect. 4.1), we have
\begin{gather}
A_{\left( \varphi (g),\widehat{\varepsilon }(g)\right) ,DSR4}(x^{5})=
\nonumber \\
\medskip  \nonumber \\
=\left(
\begin{array}{cccc}
0 & 0 & 0 & 0 \\
0 & 0 & \theta ^{3}(g)b_{1}^{-2}(x^{5}) & -\theta ^{2}(g)b_{1}^{-2}(x^{5})
\\
0 & -\theta ^{3}(g)b_{2}^{-2}(x^{5}) & 0 & \theta ^{1}(g)b_{2}^{-2}(x^{5})
\\
0 & \theta ^{2}(g)b_{3}^{-2}(x^{5}) & -\theta ^{1}(g)b_{3}^{-2}(x^{5}) & 0
\end{array}
\right) ,
\end{gather}
and therefore the corresponding finite form is given by the matrix
exponential
\begin{gather}
\exp \left( A_{\left( \varphi (g),\widehat{\varepsilon }(g)\right)
,DSR4}(x^{5})\right) =  \nonumber \\
\medskip  \nonumber \\
=\exp \left(
\begin{array}{cccc}
0 & 0 & 0 & 0 \\
0 & 0 & \theta ^{3}(g)b_{1}^{-2}(x^{5}) & -\theta ^{2}(g)b_{1}^{-2}(x^{5})
\\
0 & -\theta ^{3}(g)b_{2}^{-2}(x^{5}) & 0 & \theta ^{1}(g)b_{2}^{-2}(x^{5})
\\
0 & \theta ^{2}(g)b_{3}^{-2}(x^{5}) & -\theta ^{1}(g)b_{3}^{-2}(x^{5}) & 0
\end{array}
\right) .
\end{gather}

It is easy to see by direct calculation that the following recursive
relations hold for the powers of the matrix $A_{\left( \varphi (g),\widehat{%
\varepsilon }(g)\right) ,DSR4}(x^{5})$:
\begin{eqnarray}
&&\left\{
\begin{array}{l}
1)\left( A_{\left( \varphi (g),\widehat{\varepsilon }(g)\right)
,DSR4}(x^{5})\right) ^{0}=1_{4-d.}\medskip \medskip \bigskip ; \\
\\
2)\left( A_{\left( \varphi (g),\widehat{\varepsilon }(g)\right)
,DSR4}(x^{5})\right) ^{2n}=\bigskip \\
=\left( -1\right) ^{n-1}\left| {\bf \tilde{\theta}}(g,x^{5})\right|
^{2(n-1)}\left( A_{\left( \varphi (g),\widehat{\varepsilon }(g)\right)
,DSR4}(x^{5})\right) ^{2}=\medskip \medskip \\
=-\left| {\bf \tilde{\theta}}(g,x^{5})\right| ^{-2}\left( -1\right)
^{n}\left| {\bf \tilde{\theta}}(g,x^{5})\right| ^{2n}\left( A_{\left(
\varphi (g),\widehat{\varepsilon }(g)\right) ,DSR4}(x^{5})\right) ^{2},\text{
}n\in N; \\
\medskip \medskip \bigskip \\
3)\left( A_{\left( \varphi (g),\widehat{\varepsilon }(g)\right)
,DSR4}(x^{5})\right) ^{2n+1}=\bigskip \\
=\left( -1\right) ^{n}\left| {\bf \tilde{\theta}}(g,x^{5})\right|
^{2n}A_{\left( \varphi (g),\widehat{\varepsilon }(g)\right) ,DSR4}(x^{5}),%
\text{ }n\in N\cup \left\{ 0\right\} \medskip \medskip .
\end{array}
\right.  \nonumber \\
&&
\end{eqnarray}
By comparing relations (92) with the corresponding ''undeformed'' ones valid
in SR, it can be immediately seen that the {\em (local) generalizing
''anisotropizing deforming''} transition SR$\rightarrow $DSR4 implies the
loss of symmetry of all the powers of $A_{\left( \varphi (g),\widehat{%
\varepsilon }(g)\right) ,DSR4}(x^{5})$ (this is due to the loss of symmetry
of the 4-d. representation of the infinitesimal generators of $%
SO(3,1)_{DEF.} $: see Ref. [1]).

Then, it is possible to evaluate the exponential of $A_{\left( \varphi (g),%
\widehat{\varepsilon }(g)\right) ,DSR4}(x^{5})$:
\begin{gather}
\exp (A_{\left( \varphi (g),\widehat{\varepsilon }(g)\right)
,DSR4}(x^{5}))=\sum_{n=0}^{\infty }\frac{1}{n!}\left( A_{\left( \varphi (g),%
\widehat{\varepsilon }(g)\right) ,DSR4}(x^{5})\right) ^{n}=  \nonumber \\
\medskip  \nonumber \\
\nonumber \\
=1_{4-d.}+\sum_{n=1}^{\infty }\frac{1}{\left( 2n\right) !}\left( A_{\left(
\varphi (g),\widehat{\varepsilon }(g)\right) ,DSR4}(x^{5})\right)
^{2n}+\medskip  \nonumber \\
\nonumber \\
+\sum_{n=0}^{\infty }\frac{1}{\left( 2n+1\right) !}\left( A_{\left( \varphi
(g),\widehat{\varepsilon }(g)\right) ,DSR4}(x^{5})\right) ^{2n+1}=  \nonumber
\\
\nonumber \\
\medskip  \nonumber \\
=1_{4-d.}-\left| {\bf \tilde{\theta}}(g,x^{5})\right| ^{-2}\left( A_{\left(
\varphi (g),\widehat{\varepsilon }(g)\right) ,DSR4}(x^{5})\right)
^{2}\sum_{n=1}^{\infty }\left( -1\right) ^{n}\frac{\left| {\bf \tilde{\theta}%
}(g,x^{5})\right| ^{2n}}{\left( 2n\right) !}+  \nonumber \\
\medskip  \nonumber \\
+A_{\left( \varphi (g),\widehat{\varepsilon }(g)\right)
,DSR4}(x^{5})\sum_{n=0}^{\infty }\left( -1\right) ^{n}\frac{\left| {\bf
\tilde{\theta}}(g,x^{5})\right| ^{2n}}{\left( 2n+1\right) !}=\medskip
\nonumber \\
\nonumber \\
=1_{4-d.}-\left| {\bf \tilde{\theta}}(g,x^{5})\right| ^{-2}\left( A_{\left(
\varphi (g),\widehat{\varepsilon }(g)\right) ,DSR4}(x^{5})\right)
^{2}\sum_{n=1}^{\infty }\left( -1\right) ^{n}\frac{\left| {\bf \tilde{\theta}%
}(g,x^{5})\right| ^{2n}}{\left( 2n\right) !}+  \nonumber \\
\medskip  \nonumber \\
+\left| {\bf \tilde{\theta}}(g,x^{5})\right| ^{-1}A_{\left( \varphi (g),%
\widehat{\varepsilon }(g)\right) ,DSR4}(x^{5})\sum_{n=0}^{\infty }\left(
-1\right) ^{n}\frac{\left| {\bf \tilde{\theta}}(g,x^{5})\right| ^{2n+1}}{%
\left( 2n+1\right) !}=  \nonumber \\
\medskip  \nonumber \\
\nonumber \\
=1_{4-d.}-\left| {\bf \tilde{\theta}}(g,x^{5})\right| ^{-2}(\cos \left| {\bf
\tilde{\theta}}(g,x^{5})\right| -1)\left( A_{\left( \varphi (g),\widehat{%
\varepsilon }(g)\right) ,DSR4}(x^{5})\right) ^{2}+  \nonumber \\
\medskip  \nonumber \\
+\left| {\bf \tilde{\theta}}(g,x^{5})\right| ^{-1}\sin \left| {\bf \tilde{%
\theta}}(g,x^{5})\right| A_{\left( \varphi (g),\widehat{\varepsilon }%
(g)\right) ,DSR4}(x^{5})\medskip .
\end{gather}

In conclusion, we have, for the finite, deformed (clockwise) true rotation
by an angle $\varphi (g)$ about a generic axis $\widehat{\varepsilon }(g)$
of $\widetilde{E_{3}}(x^{5})\left( \subset \widetilde{M_{4}}(x^{5})\right) $%
:
\begin{gather}
\left(
\begin{array}{c}
\left( x^{\prime }\right) _{(g),DSR4}^{0}(\left\{ x\right\} _{m.},x^{5}) \\
\left( x^{\prime }\right) _{(g),DSR4}^{1}(\left\{ x\right\} _{m.},x^{5}) \\
\left( x^{\prime }\right) _{(g),DSR4}^{2}(\left\{ x\right\} _{m.},x^{5}) \\
\left( x^{\prime }\right) _{(g),DSR4}^{3}(\left\{ x\right\} _{m.},x^{5})
\end{array}
\right) =\left(
\begin{array}{c}
\left( x^{0}\right) _{(g),DSR4}^{^{\prime }}(\left\{ x\right\} _{m.},x^{5})
\\
\left( x^{1}\right) _{(g),DSR4}^{^{\prime }}(\left\{ x\right\} _{m.},x^{5})
\\
\left( x^{2}\right) _{(g),DSR4}^{^{\prime }}(\left\{ x\right\} _{m.},x^{5})
\\
\left( x^{3}\right) _{(g),DSR4}^{^{\prime }}(\left\{ x\right\} _{m.},x^{5})
\end{array}
\right) =  \nonumber \\
\nonumber \\
=\left( \exp (A_{\left( \varphi (g),\widehat{\varepsilon }(g)\right)
,DSR4}(x^{5}))\right) \left(
\begin{array}{c}
x^{0} \\
x^{1} \\
x^{2} \\
x^{3}
\end{array}
\right) ,
\end{gather}
with explicit components (where $\bullet $ indicates the usual algebraic
multiplication)
\begin{gather}
\left( x^{\prime }\right) _{(g),DSR4}^{0}(\left\{ x\right\}
_{m.},x^{5})=\left( x^{0}\right) _{(g),DSR4}^{^{\prime }}(\left\{ x\right\}
_{m.},x^{5})=  \nonumber \\
\nonumber \\
=\sum_{\mu =0}^{3}\left( \exp (A_{\left( \varphi (g),\widehat{\varepsilon }%
(g)\right) ,DSR4}(x^{5}))\right) _{~~\mu }^{0}x^{\mu }=x^{0};
\end{gather}
\begin{gather}
\left( x^{\prime }\right) _{(g),DSR4}^{1}(\left\{ x\right\}
_{m.},x^{5})=\left( x^{1}\right) _{(g),DSR4}^{^{\prime }}(\left\{ x\right\}
_{m.},x^{5})=\sum_{\mu =0}^{3}\left( \exp (A_{\left( \varphi (g),\widehat{%
\varepsilon }(g)\right) ,DSR4}(x^{5}))\right) _{~~\mu }^{1}x^{\mu }=
\nonumber \\
\nonumber \\
=\left[ 1+\frac{(\cos \left| {\bf \tilde{\theta}}(g,x^{5})\right| -1)\left(
\left| {\bf \tilde{\theta}}(g,x^{5})\right| ^{2}-\left( \theta
^{1}(g)\right) ^{2}b_{2}^{-2}(x^{5})b_{3}^{-2}(x^{5})\right) }{\left| {\bf
\tilde{\theta}}(g,x^{5})\right| ^{2}}\right] x^{1}+  \nonumber \\
\medskip  \nonumber \\
+\frac{\theta ^{3}(g)b_{1}^{-2}(x^{5})\left| {\bf \tilde{\theta}}%
(g,x^{5})\right| \sin \left| {\bf \tilde{\theta}}(g,x^{5})\right| }{\left|
{\bf \tilde{\theta}}(g,x^{5})\right| ^{2}}x^{2}+\medskip \medskip  \nonumber
\\
\nonumber \\
-\frac{(\cos \left| {\bf \tilde{\theta}}(g,x^{5})\right| -1)\theta
^{1}(g)\theta ^{2}(g)b_{1}^{-2}(x^{5})b_{3}^{-2}(x^{5})}{\left| {\bf \tilde{%
\theta}}(g,x^{5})\right| ^{2}}x^{2}+\medskip  \nonumber \\
\nonumber \\
-\frac{\theta ^{2}(g)b_{1}^{-2}(x^{5})\left| {\bf \tilde{\theta}}%
(g,x^{5})\right| \sin \left| {\bf \tilde{\theta}}(g,x^{5})\right| }{\left|
{\bf \tilde{\theta}}(g,x^{5})\right| ^{2}}x^{3}+\medskip  \nonumber \\
\nonumber \\
+\frac{(\cos \left| {\bf \tilde{\theta}}(g,x^{5})\right| -1)\theta
^{1}(g)\theta ^{3}(g)b_{1}^{-2}(x^{5})b_{2}^{-2}(x^{5})}{\left| {\bf \tilde{%
\theta}}(g,x^{5})\right| ^{2}}x^{3}=\medskip \medskip \medskip  \nonumber \\
\nonumber \\
=\frac{1}{\left| {\bf \tilde{\theta}}(g,x^{5})\right| ^{2}}\left\{ \left[
\left| {\bf \tilde{\theta}}(g,x^{5})\right| ^{2}+\right. \right. \medskip
\medskip  \nonumber \\
\nonumber \\
\left. +(\cos \left| {\bf \tilde{\theta}}(g,x^{5})\right| -1)\left( \left|
{\bf \tilde{\theta}}(g,x^{5})\right| ^{2}-\left( \theta ^{1}(g)\right)
^{2}b_{2}^{-2}(x^{5})b_{3}^{-2}(x^{5})\right) \right] x^{1}+\medskip
\medskip \medskip  \nonumber \\
\nonumber \\
+\left[ \theta ^{3}(g)b_{1}^{-2}(x^{5})\left| {\bf \tilde{\theta}}%
(g,x^{5})\right| \sin \left| {\bf \tilde{\theta}}(g,x^{5})\right| +\right.
\medskip \medskip  \nonumber \\
\nonumber \\
\left. -(\cos \left| {\bf \tilde{\theta}}(g,x^{5})\right| -1)\theta
^{1}(g)\theta ^{2}(g)b_{1}^{-2}(x^{5})b_{3}^{-2}(x^{5})\right]
x^{2}+\medskip \medskip \medskip  \nonumber \\
\nonumber \\
-\left[ \theta ^{2}(g)b_{1}^{-2}(x^{5})\left| {\bf \tilde{\theta}}%
(g,x^{5})\right| \sin \left| {\bf \tilde{\theta}}(g,x^{5})\right| +\right.
\medskip \medskip  \nonumber \\
\nonumber \\
\left. \left. +(\cos \left| {\bf \tilde{\theta}}(g,x^{5})\right| -1)\theta
^{1}(g)\theta ^{3}(g)b_{1}^{-2}(x^{5})b_{2}^{-2}(x^{5})\right] x^{3}\right\}
\medskip \medskip \medskip ;
\end{gather}
\begin{gather}
\left( x^{\prime }\right) _{(g),DSR4}^{2}(\left\{ x\right\}
_{m.},x^{5})=\left( x^{2}\right) _{(g),DSR4}^{^{\prime }}(\left\{ x\right\}
_{m.},x^{5})=\sum_{\mu =0}^{3}\left( \exp (A_{\left( \varphi (g),\widehat{%
\varepsilon }(g)\right) ,DSR4}(x^{5}))\right) _{~~\mu }^{2}x^{\mu }=
\nonumber \\
\medskip \medskip  \nonumber \\
=-\frac{\theta ^{3}(g)b_{2}^{-2}(x^{5})\left| {\bf \tilde{\theta}}%
(g,x^{5})\right| \sin \left| {\bf \tilde{\theta}}(g,x^{5})\right| }{\left|
{\bf \tilde{\theta}}(g,x^{5})\right| ^{2}}x^{1}+\medskip \medskip  \nonumber
\\
\nonumber \\
+\frac{(\cos \left| {\bf \tilde{\theta}}(g,x^{5})\right| -1)\theta
^{1}(g)\theta ^{2}(g)b_{2}^{-2}(x^{5})b_{3}^{-2}(x^{5})}{\left| {\bf \tilde{%
\theta}}(g,x^{5})\right| ^{2}}x^{1}+\medskip  \nonumber \\
\nonumber \\
+\left[ 1+\frac{(\cos \left| {\bf \tilde{\theta}}(g,x^{5})\right| -1)\left(
\left| {\bf \tilde{\theta}}(g,x^{5})\right| ^{2}-\left( \theta
^{2}(g)\right) ^{2}b_{1}^{-2}(x^{5})b_{3}^{-2}(x^{5})\right) }{\left| {\bf
\tilde{\theta}}(g,x^{5})\right| ^{2}}\right] x^{2}+\medskip  \nonumber \\
\nonumber \\
+\frac{\theta ^{1}(g)b_{2}^{-2}(x^{5})\left| {\bf \tilde{\theta}}%
(g,x^{5})\right| \sin \left| {\bf \tilde{\theta}}(g,x^{5})\right| }{\left|
{\bf \tilde{\theta}}(g,x^{5})\right| ^{2}}x^{3}+\medskip  \nonumber \\
\nonumber \\
-\frac{(\cos \left| {\bf \tilde{\theta}}(g,x^{5})\right| -1)\theta
^{2}(g)\theta ^{3}(g)b_{1}^{-2}(x^{5})b_{2}^{-2}(x^{5})}{\left| {\bf \tilde{%
\theta}}(g,x^{5})\right| ^{2}}x^{3}=\medskip \medskip \medskip  \nonumber \\
\nonumber \\
=\frac{1}{\left| {\bf \tilde{\theta}}(g,x^{5})\right| ^{2}}\left\{ -\left[
\theta ^{3}(g)b_{2}^{-2}(x^{5})\left| {\bf \tilde{\theta}}(g,x^{5})\right|
\sin \left| {\bf \tilde{\theta}}(g,x^{5})\right| +\right. \right. \medskip
\medskip  \nonumber \\
\nonumber \\
\left. +(\cos \left| {\bf \tilde{\theta}}(g,x^{5})\right| -1)\theta
^{1}(g)\theta ^{2}(g)b_{2}^{-2}(x^{5})b_{3}^{-2}(x^{5})\right]
x^{1}+\medskip \medskip \medskip  \nonumber \\
\nonumber \\
+\left[ \left| {\bf \tilde{\theta}}(g,x^{5})\right| ^{2}+(\cos \left| {\bf
\tilde{\theta}}(g,x^{5})\right| -1)\bullet \right. \medskip \medskip
\nonumber \\
\nonumber \\
\left. \bullet \left( \left| {\bf \tilde{\theta}}(g,x^{5})\right|
^{2}-\left( \theta ^{2}(g)\right)
^{2}b_{1}^{-2}(x^{5})b_{3}^{-2}(x^{5})\right) \right] x^{2}+\medskip
\medskip \medskip  \nonumber \\
\nonumber \\
+\left[ \theta ^{1}(g)b_{2}^{-2}(x^{5})\left| {\bf \tilde{\theta}}%
(g,x^{5})\right| \sin \left| {\bf \tilde{\theta}}(g,x^{5})\right| +\right.
\medskip \medskip  \nonumber \\
\nonumber \\
\left. \left. -(\cos \left| {\bf \tilde{\theta}}(g,x^{5})\right| -1)\theta
^{2}(g)\theta ^{3}(g)b_{1}^{-2}(x^{5})b_{2}^{-2}(x^{5})\right] x^{3}\medskip
\right\} \medskip \medskip ;
\end{gather}

\begin{gather}
\left( x^{\prime }\right) _{(g),DSR4}^{3}(\left\{ x\right\}
_{m.},x^{5})=\left( x^{3}\right) _{(g),DSR4}^{^{\prime }}(\left\{ x\right\}
_{m.},x^{5})=\sum_{\mu =0}^{3}\left( \exp (A_{\left( \varphi (g),\widehat{%
\varepsilon }(g)\right) ,DSR4}(x^{5}))\right) _{~~\mu }^{3}x^{\mu }=
\nonumber \\
\medskip \medskip  \nonumber \\
=\frac{\theta ^{2}(g)b_{3}^{-2}(x^{5})\left| {\bf \tilde{\theta}}%
(g,x^{5})\right| \sin \left| {\bf \tilde{\theta}}(g,x^{5})\right| }{\left|
{\bf \tilde{\theta}}(g,x^{5})\right| ^{2}}x^{1}+\medskip \medskip  \nonumber
\\
\nonumber \\
-\frac{(\cos \left| {\bf \tilde{\theta}}(g,x^{5})\right| -1)\theta
^{1}(g)\theta ^{3}(g)b_{2}^{-2}(x^{5})b_{3}^{-2}(x^{5})}{\left| {\bf \tilde{%
\theta}}(g,x^{5})\right| ^{2}}x^{1}+\medskip  \nonumber \\
\nonumber \\
-\frac{\theta ^{1}(g)b_{3}^{-2}(x^{5})\left| {\bf \tilde{\theta}}%
(g,x^{5})\right| \sin \left| {\bf \tilde{\theta}}(g,x^{5})\right| }{\left|
{\bf \tilde{\theta}}(g,x^{5})\right| ^{2}}x^{2}+  \nonumber \\
\nonumber \\
+\frac{(\cos \left| {\bf \tilde{\theta}}(g,x^{5})\right| -1)\theta
^{2}(g)\theta ^{3}(g)b_{1}^{-2}(x^{5})b_{3}^{-2}(x^{5})}{\left| {\bf \tilde{%
\theta}}(g,x^{5})\right| ^{2}}x^{2}+\medskip  \nonumber \\
\nonumber \\
+\left[ 1+\frac{(\cos \left| {\bf \tilde{\theta}}(g,x^{5})\right| -1)\left(
\left| {\bf \tilde{\theta}}(g,x^{5})\right| ^{2}-\left( \theta
^{3}(g)\right) ^{2}b_{1}^{-2}(x^{5})b_{2}^{-2}(x^{5})\right) }{\left| {\bf
\tilde{\theta}}(g,x^{5})\right| ^{2}}\right] x^{3}=  \nonumber \\
\medskip \medskip \medskip  \nonumber \\
=\frac{1}{\left| {\bf \tilde{\theta}}(g,x^{5})\right| ^{2}}\left\{ \left[
\theta ^{2}(g)b_{3}^{-2}(x^{5})\left| {\bf \tilde{\theta}}(g,x^{5})\right|
\sin \left| {\bf \tilde{\theta}}(g,x^{5})\right| +\right. \right. \medskip
\medskip  \nonumber \\
\nonumber \\
\left. -(\cos \left| {\bf \tilde{\theta}}(g,x^{5})\right| -1)\theta
^{1}(g)\theta ^{3}(g)b_{2}^{-2}(x^{5})b_{3}^{-2}(x^{5})\right]
x^{1}+\medskip \medskip \medskip  \nonumber \\
\nonumber \\
-\left[ \theta ^{1}(g)b_{3}^{-2}(x^{5})\left| {\bf \tilde{\theta}}%
(g,x^{5})\right| \sin \left| {\bf \tilde{\theta}}(g,x^{5})\right| +\right.
\medskip \medskip  \nonumber \\
\nonumber \\
\left. +(\cos \left| {\bf \tilde{\theta}}(g,x^{5})\right| -1)\theta
^{2}(g)\theta ^{3}(g)b_{1}^{-2}(x^{5})b_{3}^{-2}(x^{5})\right]
x^{2}+\medskip \medskip \medskip  \nonumber \\
\nonumber \\
+\left[ \left| {\bf \tilde{\theta}}(g,x^{5})\right| ^{2}+(\cos \left| {\bf
\tilde{\theta}}(g,x^{5})\right| -1)\bullet \right. \medskip \medskip
\nonumber \\
\nonumber \\
\left. \left. \bullet \left( \left| {\bf \tilde{\theta}}(g,x^{5})\right|
^{2}-\left( \theta ^{3}(g)\right)
^{2}b_{1}^{-2}(x^{5})b_{2}^{-2}(x^{5})\right) \right] x^{3}\right\} \medskip
\medskip \medskip .
\end{gather}

\section{FINITE 3-d. DEFORMED BOOSTS\ IN\ A\ GENERIC\ DIRECTION\ IN DSR4}

\subsection{\protect\bigskip Parametric Decomposition}

We want now to derive the finite, deformed pseudorotations (or ''boosts'')
with generic dimensionless parameter (''rapidity'') $\rho (g)$ along a
generic direction $\widehat{\varepsilon }(g)$ in the physical 3-d. space $%
\widetilde{E_{3}}(x^{5})\subset \widetilde{M_{4}}(x^{5})$, by exploiting the
form of the infinitesimal generators of the DSR4 chronotopical group $%
SO(3,1)_{DEF.}$ obtained in Ref. [1].

We can proceed in a similar way to the case of finite deformed true
rotations treated in the previous Section 5. Thus we have the following {\em %
(not unique)} axial-parametric decomposition:
\begin{equation}
\left( \widehat{\varepsilon }(g),\rho (g)\right) \rightarrow (\widehat{x^{1}}%
,\zeta ^{1}(g))(\widehat{x^{2}},\zeta ^{2}(g))(\widehat{x^{3}},\zeta
^{3}(g)),
\end{equation}
whence in DSR4 we obtain ($\cdot $ denotes the Euclidean scalar product):
\begin{gather}
-\rho (g)\widehat{\varepsilon }(g)\cdot {\bf K}_{DSR4}(x^{5})\rightarrow
\nonumber \\
\nonumber \\
\rightarrow -\zeta ^{1}(g)K_{DSR4}^{1}(x^{5})-\zeta
^{2}(g)K_{DSR4}^{2}(x^{5})-\zeta ^{3}(g)K_{DSR4}^{3}(x^{5})=  \nonumber \\
\medskip  \nonumber \\
=-\sum_{i=1}^{3}\zeta ^{i}(g)K_{DSR4}^{i}(x^{5})\medskip
\end{gather}
(infinitesimal level , any composition order);
\begin{gather}
\exp \left( -\rho (g)\widehat{\varepsilon }(g)\cdot {\bf K}%
_{DSR4}(x^{5})\right) \rightarrow  \nonumber \\
\\
\rightarrow \exp \left( -\zeta ^{1}(g)K_{DSR4}^{1}(x^{5})\right) \times \exp
\left( -\zeta ^{2}(g)K_{DSR4}^{2}(x^{5})\right) \times \exp \left( -\zeta
^{3}(g)K_{DSR4}^{3}(x^{5})\right) \medskip  \nonumber
\end{gather}
(finite level, fixed composition order).

Denoting by $B_{\left( \rho (g),\widehat{\varepsilon }(g)\right)
,DSR4}(x^{5})$ the 4$\times $4 matrix corresponding to an infinitesimal
boost with (infinitesimal) rapidity $\rho (g)$ about the axis $\widehat{%
\varepsilon }(g)$ (matrix belonging a 4-d. representation of the deformed
Lorentz algebra $su(2)_{DEF.}\times su(2)_{DEF.}$), we have to go through
the following steps:

\begin{gather}
B_{\left( \rho (g),\widehat{\varepsilon }(g)\right) ,DSR4}(x^{5})\equiv
-\rho (g)\widehat{\varepsilon }(g)\cdot {\bf K}_{DSR4}(x^{5})\rightarrow
\nonumber \\
\nonumber \\
\rightarrow \left( -\zeta ^{1}(g)K_{DSR4}^{1}(x^{5})-\zeta
^{2}(g)K_{DSR4}^{2}(x^{5})-\zeta ^{3}(g)K_{DSR4}^{3}(x^{5})\right)
\rightarrow  \nonumber \\
\medskip  \nonumber \\
\rightarrow \exp (B_{\left( \rho (g),\widehat{\varepsilon }(g)\right)
,DSR4}(x^{5}))=\medskip  \nonumber \\
\nonumber \\
=\exp \left( -\zeta ^{1}(g)K_{DSR4}^{1}(x^{5})-\zeta
^{2}(g)K_{DSR4}^{2}(x^{5})-\zeta ^{3}(g)K_{DSR4}^{3}(x^{5})\right) =
\nonumber \\
\medskip  \nonumber \\
=\sum_{n=0}^{\infty }\frac{1}{n!}\left( -\zeta
^{1}(g)K_{DSR4}^{1}(x^{5})-\zeta ^{2}(g)K_{DSR4}^{2}(x^{5})-\zeta
^{3}(g)K_{DSR4}^{3}(x^{5})\right) ^{n}\medskip ,
\end{gather}
where $\exp (B_{\left( \rho (g),\widehat{\varepsilon }(g)\right)
,DSR4}(x^{5}))$ is the 4$\times $4 matrix corresponding to a finite
pseudorotation with finite rapidity $\varphi (g)$ along the axis $\widehat{%
\varepsilon }(g)$, belonging to a 4-d. representation of $SO(3,1)_{DEF.}$,
of course.

On account of the explicit form of ${\bf K}_{DSR4}(x^{5})$ and of the
deformed boost generators (see Subsect. 4.1), we have
\begin{gather}
B_{\left( \rho (g),\widehat{\varepsilon }(g)\right) ,DSR4}(x^{5})=\medskip
\nonumber \\
\nonumber \\
=\left(
\begin{array}{cccc}
0 & -\zeta ^{1}(g)b_{0}^{-2}(x^{5}) & -\zeta ^{2}(g)b_{0}^{-2}(x^{5}) &
-\zeta ^{3}(g)b_{0}^{-2}(x^{5}) \\
-\zeta ^{1}(g)b_{1}^{-2}(x^{5}) & 0 & 0 & 0 \\
-\zeta ^{2}(g)b_{2}^{-2}(x^{5}) & 0 & 0 & 0 \\
-\zeta ^{3}(g)b_{3}^{-2}(x^{5}) & 0 & 0 & 0
\end{array}
\right) ,  \nonumber \\
\end{gather}
and therefore the corresponding finite form is given by the matrix
exponential
\begin{gather}
\exp \left( B_{\left( \rho (g),\widehat{\varepsilon }(g)\right)
,DSR4}(x^{5})\right) =\medskip   \nonumber \\
\nonumber \\
=\exp \left(
\begin{array}{cccc}
0 & -\zeta ^{1}(g)b_{0}^{-2}(x^{5}) & -\zeta ^{2}(g)b_{0}^{-2}(x^{5}) &
-\zeta ^{3}(g)b_{0}^{-2}(x^{5}) \\
-\zeta ^{1}(g)b_{1}^{-2}(x^{5}) & 0 & 0 & 0 \\
-\zeta ^{2}(g)b_{2}^{-2}(x^{5}) & 0 & 0 & 0 \\
-\zeta ^{3}(g)b_{3}^{-2}(x^{5}) & 0 & 0 & 0
\end{array}
\right) .  \nonumber \\
\end{gather}
Therefore, in matrix form, the deformed boost transformation reads
\begin{gather}
\left(
\begin{array}{c}
\left( x^{\prime }\right) _{(g),DSR4}^{0}(\left\{ x\right\} _{m.},x^{5}) \\
\left( x^{\prime }\right) _{(g),DSR4}^{1}(\left\{ x\right\} _{m.},x^{5}) \\
\left( x^{\prime }\right) _{(g),DSR4}^{2}(\left\{ x\right\} _{m.},x^{5}) \\
\left( x^{\prime }\right) _{(g),DSR4}^{3}(\left\{ x\right\} _{m.},x^{5})
\end{array}
\right) =\left(
\begin{array}{c}
\left( x^{0}\right) _{(g),DSR4}^{^{\prime }}(\left\{ x\right\} _{m.},x^{5})
\\
\left( x^{1}\right) _{(g),DSR4}^{^{\prime }}(\left\{ x\right\} _{m.},x^{5})
\\
\left( x^{2}\right) _{(g),DSR4}^{^{\prime }}(\left\{ x\right\} _{m.},x^{5})
\\
\left( x^{3}\right) _{(g),DSR4}^{^{\prime }}(\left\{ x\right\} _{m.},x^{5})
\end{array}
\right) =  \nonumber \\
\nonumber \\
=\exp \left( B_{\left( \rho (g),\widehat{\varepsilon }(g)\right)
,DSR4}(x^{5})\right) \left(
\begin{array}{c}
x^{0} \\
x^{1} \\
x^{2} \\
x^{3}
\end{array}
\right) \Leftrightarrow   \nonumber \\
\nonumber \\
\nonumber \\
\Leftrightarrow \left(
\begin{array}{c}
\left( x^{\prime }\right) _{(g),DSR4}^{0}(\left\{ x\right\} _{m.},x^{5}) \\
\left( x^{\prime }\right) _{(g),DSR4}^{1}(\left\{ x\right\} _{m.},x^{5}) \\
\left( x^{\prime }\right) _{(g),DSR4}^{2}(\left\{ x\right\} _{m.},x^{5}) \\
\left( x^{\prime }\right) _{(g),DSR4}^{3}(\left\{ x\right\} _{m.},x^{5})
\end{array}
\right) =\left(
\begin{array}{c}
\left( x^{0}\right) _{(g),DSR4}^{^{\prime }}(\left\{ x\right\} _{m.},x^{5})
\\
\left( x^{1}\right) _{(g),DSR4}^{^{\prime }}(\left\{ x\right\} _{m.},x^{5})
\\
\left( x^{2}\right) _{(g),DSR4}^{^{\prime }}(\left\{ x\right\} _{m.},x^{5})
\\
\left( x^{3}\right) _{(g),DSR4}^{^{\prime }}(\left\{ x\right\} _{m.},x^{5})
\end{array}
\right) =  \nonumber \\
\nonumber \\
\nonumber \\
=\exp \left(
\begin{array}{cccc}
0 & -\zeta ^{1}(g)b_{0}^{-2}(x^{5}) & -\zeta ^{2}(g)b_{0}^{-2}(x^{5}) &
-\zeta ^{3}(g)b_{0}^{-2}(x^{5}) \\
-\zeta ^{1}(g)b_{1}^{-2}(x^{5}) & 0 & 0 & 0 \\
-\zeta ^{2}(g)b_{2}^{-2}(x^{5}) & 0 & 0 & 0 \\
-\zeta ^{3}(g)b_{3}^{-2}(x^{5}) & 0 & 0 & 0
\end{array}
\right) \left(
\begin{array}{c}
x^{0} \\
x^{1} \\
x^{2} \\
x^{3}
\end{array}
\right) .  \nonumber \\
\\
\nonumber
\end{gather}

\subsection{Deformed Boost from Velocity Decomposition}

As noticed in Sect. 5, all the possible procedures applicable in these
frameworks are {\em equivalent} (although they may yield different formal
results). Thus, in order to obtain the general form of a finite, deformed
boost with generic dimensionless parameter (''rapidity'') $\rho (g)$ along a
generic direction $\widehat{\varepsilon }(g)$ in $\widetilde{E_{3}}%
(x^{5})\subset \widetilde{M_{4}}(x^{5})$, let us exploit another approach
instead of explicitly calculating $\exp \left( B_{\left( \rho (g),\widehat{%
\varepsilon }(g)\right) ,DSR4}(x^{5})\right) $.

Consider a generic finite, deformed boost with generic velocity ${\bf v}%
(g)=v(g)\widehat{v}(g)\equiv v(g)\widehat{\varepsilon }(g)$ along a generic
direction $\widehat{\varepsilon }(g)$ in $\widetilde{E_{3}}(x^{5})\subset
\widetilde{M_{4}}(x^{5})$. Let us decompose the 3-vector ${\bf x}${\bf \ }in
two components ${\bf x}_{\parallel }(g)$ and ${\bf x}_{\perp }(g)$,
respectively parallel and orthogonal to ${\bf v}(g)$. Here, {\em %
''parallelism''} and {\em ''orthogonality''} are to be meant in the deformed
3-d. space $\widetilde{E_{3}}(x^{5})\subset \widetilde{M_{4}}(x^{5})$
(namely according to the deformed scalar product $\ast $ associated to the
3-d. metric tensor $-g_{ij,DSR4}(x^{5})\stackrel{\text{{\footnotesize ESC off%
}}}{=}b_{i}^{2}(x^{5})\delta _{ij}$). We have

\begin{gather}
{\bf x}_{\parallel }(g)\equiv \widehat{v}(g)(\widehat{v}(g)\ast {\bf x})=%
\frac{{\bf v}(g)}{\left| {\bf v}(g)\right| _{\ast }^{2}}({\bf v}(g)\ast {\bf %
x})=\frac{{\bf v}(g)}{{\bf v}(g)\ast {\bf v}(g)}({\bf v}(g)\ast {\bf x})=
\nonumber \\
\nonumber \\
=\frac{\sum_{i=1}^{3}b_{i}^{2}(x^{5})v^{i}(g)x^{i}}{%
\sum_{i=1}^{3}b_{i}^{2}(x^{5})\left( v^{i}(g)\right) ^{2}}{\bf v}(g)%
\stackrel{\text{{\footnotesize in DSR4:}}{\bf \tilde{\beta}}(g)\equiv \left(
\frac{{\bf v(}g{\bf )}}{{\bf u}}\right) \neq \frac{{\bf v}(g)}{u}}{\neq }%
\widehat{\widetilde{\beta }}(g)(\widehat{\widetilde{\beta }}(g)\ast {\bf x})=
\nonumber \\
\nonumber \\
=\frac{{\bf \tilde{\beta}}(g)}{\left| {\bf \tilde{\beta}}(g)\right| _{\ast
}^{2}}({\bf \tilde{\beta}}(g)\ast {\bf x})=\frac{{\bf \tilde{\beta}}(g)}{%
{\bf \tilde{\beta}}(g)\ast {\bf \tilde{\beta}}(g)}({\bf \tilde{\beta}}%
(g)\ast {\bf x})=  \nonumber \\
\nonumber \\
=\frac{\sum_{i=1}^{3}b_{i}^{2}(x^{5})\widetilde{\beta }^{i}(g)x^{i}}{%
\sum_{i=1}^{3}b_{i}^{2}(x^{5})\left( \widetilde{\beta }^{i}(g)\right) ^{2}}%
{\bf \tilde{\beta}}(g)\medskip ;  \nonumber \\
\\
\nonumber
\end{gather}

\begin{gather}
x_{\parallel }^{i}(g)\equiv \frac{\sum_{k=1}^{3}b_{k}^{2}(x^{5})v^{k}(g)x^{k}%
}{\sum_{k=1}^{3}b_{k}^{2}(x^{5})\left( v^{k}(g)\right) ^{2}}v^{i}\neq
\nonumber \\
\nonumber \\
\stackrel{\text{{\footnotesize in DSR4:}}{\bf \tilde{\beta}}(g)\equiv \left(
\frac{{\bf v(}g{\bf )}}{{\bf u}}\right) \neq \frac{{\bf v}(g)}{u}}{\neq }%
\frac{\sum_{k=1}^{3}b_{k}^{2}(x^{5})\widetilde{\beta }^{k}(g)x^{k}}{%
\sum_{k=1}^{3}b_{k}^{2}(x^{5})\left( \widetilde{\beta }^{k}(g)\right) ^{2}}%
\widetilde{\beta }^{i}(g),\forall i=1,2,3;  \nonumber \\
\\
\nonumber
\end{gather}

\begin{gather}
{\bf x}_{\perp }(g)\equiv {\bf x}-{\bf x}_{\parallel }(g)={\bf x}-\frac{%
\sum_{i=1}^{3}b_{i}^{2}(x^{5})v^{i}(g)x^{i}}{\sum_{i=1}^{3}b_{i}^{2}(x^{5})%
\left( v^{i}(g)\right) ^{2}}{\bf v}(g)\neq  \nonumber \\
\nonumber \\
\stackrel{\text{{\footnotesize in DSR4:}}{\bf \tilde{\beta}}(g)\equiv \left(
\frac{{\bf v(}g{\bf )}}{{\bf u}}\right) \neq \frac{{\bf v}(g)}{u}}{\neq }%
{\bf x}-\frac{\sum_{i=1}^{3}b_{i}^{2}(x^{5})\widetilde{\beta }^{i}(g)x^{i}}{%
\sum_{i=1}^{3}b_{i}^{2}(x^{5})\left( \widetilde{\beta }^{i}(g)\right) ^{2}}%
{\bf \tilde{\beta}}(g);  \nonumber \\
\\
\nonumber
\end{gather}
\begin{gather}
x_{\perp }^{i}(g)\equiv x^{i}-\frac{%
\sum_{k=1}^{3}b_{k}^{2}(x^{5})v^{k}(g)x^{k}}{\sum_{k=1}^{3}b_{k}^{2}(x^{5})%
\left( v^{k}(g)\right) ^{2}}v^{i}(g)\neq  \nonumber \\
\nonumber \\
\stackrel{\text{{\footnotesize in DSR4:}}{\bf \tilde{\beta}}(g)\equiv \left(
\frac{{\bf v(}g{\bf )}}{{\bf u}}\right) \neq \frac{{\bf v}(g)}{u}}{\neq }%
x^{i}-\frac{\sum_{k=1}^{3}b_{k}^{2}(x^{5})\widetilde{\beta }^{k}(g)x^{k}}{%
\sum_{k=1}^{3}b_{k}^{2}(x^{5})\left( \widetilde{\beta }^{k}(g)\right) ^{2}}%
\widetilde{\beta }^{i}(g)\medskip ,\forall i=1,2,3,  \nonumber \\
\\
\nonumber
\end{gather}
where
\begin{gather}
{\bf \tilde{\beta}}(g)\equiv \frac{{\bf v(}g{\bf )}}{{\bf u}}=  \nonumber \\
\nonumber \\
=\left( \frac{v^{1}(g)b_{1}(x^{5})}{cb_{0}(x^{5})}\widehat{x},\frac{%
v^{2}(g)b_{2}(x^{5})}{cb_{0}(x^{5})}\widehat{y},\frac{v^{3}(g)b_{3}(x^{5})}{%
cb_{0}(x^{5})}\widehat{z}\right) \text{ .}
\end{gather}

On account of the form of a finite, deformed boost along a coordinate axis
(see Eq. (47)), a finite, deformed boost with generic (finite) velocity $%
{\bf v}(g)$ in a generic direction $\widehat{v}(g)$ is therefore given by
(see Ref. [10]) ($\cdot $ denotes, as before, Euclidean 3-d. scalar product)
\begin{eqnarray}
&&\left\{
\begin{array}{lll}
\begin{array}{l}
{\bf x}_{\parallel }{}^{\prime }(g)
\end{array}
&
\begin{array}{l}
=
\end{array}
&
\begin{array}{c}
\widetilde{\gamma }(g)({\bf x}_{\parallel }(g)-{\bf v}(g)t) \\
\medskip
\end{array}
\\
\begin{array}{l}
{\bf x}_{\perp }^{\prime }(g)
\end{array}
&
\begin{array}{l}
=
\end{array}
&
\begin{array}{c}
\begin{array}{c}
{\bf x}_{\perp }(g)\medskip
\end{array}
\end{array}
\\
\begin{array}{l}
t^{\prime }
\end{array}
&
\begin{array}{l}
=
\end{array}
& \left\{
\begin{array}{l}
\widetilde{\gamma }(g)\left( t-\sum_{i=1}^{3}\frac{v^{i}(g)b_{i}^{2}(x^{5})}{%
c^{2}b_{0}^{2}(x^{5})}x^{i}\right) =\widetilde{\gamma }(g)\left( t-{\bf
\tilde{B}}(g)\cdot {\bf x}\right) = \\
\\
\\
=\widetilde{\gamma }\left( t-{\bf \tilde{B}}^{(\ast )}(g)\ast {\bf x}%
)\right) ,
\end{array}
\right.
\end{array}
\right.  \nonumber \\
&&
\end{eqnarray}
where
\begin{gather}
\widetilde{\gamma }(g)\equiv \left( 1-{\bf \tilde{\beta}}(g)\cdot {\bf
\tilde{\beta}}(g)\right) ^{-1/2}=\left( 1-{\bf \tilde{\beta}}^{(\ast
)}(g)\ast {\bf \tilde{\beta}}^{(\ast )}(g)\right) ^{-1/2}=  \nonumber \\
\nonumber \\
=\left( 1-\left( \frac{v^{1}(g)b_{1}(x^{5})}{cb_{0}(x^{5})}\right)
^{2}-\left( \frac{v^{2}(g)b_{2}(x^{5})}{cb_{0}(x^{5})}\right) ^{2}-\left(
\frac{v^{3}(g)b_{3}(x^{5})}{cb_{0}(x^{5})}\right) ^{2}\right) ^{-1/2};
\nonumber \\
\end{gather}

\begin{gather}
{\bf \tilde{\beta}}^{(\ast )}(g)\equiv \frac{{\bf v(}g{\bf )}}{{\bf w}}=
\nonumber \\
\nonumber \\
=\left( \frac{v^{1}(g)}{cb_{0}(x^{5})}\widehat{x},\frac{v^{2}(g)}{%
cb_{0}(x^{5})}\widehat{y},\frac{v^{3}(g)}{cb_{0}(x^{5})}\widehat{z}\right) =%
\frac{1}{cb_{0}(x^{5})}{\bf v}(g)\text{\ ;}  \nonumber \\
\end{gather}
\begin{gather}
{\bf \tilde{B}}(g)\equiv \frac{{\bf v(}g{\bf )}}{{\bf u}^{2}}=  \nonumber \\
\nonumber \\
=\left( \frac{v^{1}(g)b_{1}^{2}(x^{5})}{c^{2}b_{0}^{2}(x^{5})}\widehat{x},%
\frac{v^{2}(g)b_{2}^{2}(x^{5})}{c^{2}b_{0}^{2}(x^{5})}\widehat{y},\frac{%
v^{3}(g)b_{3}^{2}(x^{5})}{c^{2}b_{0}^{2}(x^{5})}\widehat{z}\right) ;
\nonumber \\
\end{gather}
\begin{equation}
{\bf \tilde{B}}^{(\ast )}(g)\equiv \frac{{\bf v(}g{\bf )}}{{\bf w}^{2}}=%
\frac{1}{c^{2}b_{0}^{2}(x^{5})}{\bf v}(g),
\end{equation}
and ${\bf w}$ is the (unphysical) isotropic maximal causal velocity (see
Ref. [10])
\begin{equation}
{\bf w}\equiv cb_{0}(x^{5})(\widehat{x},\widehat{y},\widehat{z}).\text{ }
\end{equation}
From the definitions (110) and (113) of ${\bf \tilde{\beta}}(g)$ and ${\bf
\tilde{\beta}}^{(\ast )}(g)$ it follows easily\footnote{%
Indeed in{\footnotesize \ }DSR4, as it may be easily verified, one has
\[
\left| {\bf \tilde{\beta}}^{(\ast )}(g)\right| _{\ast }^{2}=\left| {\bf
\tilde{\beta}}(g)\right| ^{2}{\footnotesize \ }
\]
and
\begin{gather*}
\left| {\bf w}(x^{5})\right| _{\ast
}^{2}=\sum_{k=1}^{3}b_{i}^{2}(x^{5})\left( w^{i}(x^{5})\right)
^{2}=c^{2}b_{0}^{2}(x^{5})\sum_{k=1}^{3}b_{i}^{2}(x^{5})\medskip \\
\\
\left| {\bf u}(x^{5})\right| ^{2}=\sum_{k=1}^{3}\left( u^{i}(x^{5})\right)
^{2}=c^{2}b_{0}^{2}(x^{5})\sum_{k=1}^{3}b_{i}^{-2}(x^{5})\medskip ,
\end{gather*}
so that, in general
\[
\left| {\bf w}(x^{5})\right| _{\ast }^{2}\neq \left| {\bf u}(x^{5})\right|
^{2}.
\]
\par
{}
\par
{\footnotesize \ }}
\begin{equation}
\left| {\bf v}(g)\right| _{\ast }^{2}\equiv
\sum_{k=1}^{3}b_{i}^{2}(x^{5})\left( v^{i}(g)\right)
^{2}=c^{2}b_{0}^{2}(x^{5})\left| {\bf \tilde{\beta}}(g)\right| ^{2};
\end{equation}

\begin{equation}
\left| {\bf v}(g)\right| _{\ast }^{2}\equiv
\sum_{k=1}^{3}b_{i}^{2}(x^{5})\left( v^{i}(g)\right)
^{2}=c^{2}b_{0}^{2}(x^{5})\left| {\bf \tilde{\beta}}^{(\ast )}(g)\right|
_{\ast }^{2}.
\end{equation}

In terms of the trivector ${\bf x}${\bf \ }$=$ ${\bf x}_{\parallel }(g)$ $+$
${\bf x}_{\perp }(g)$, the deformed boost (111) in a generic direction $%
\widehat{v}(g)$ reads therefore
\begin{eqnarray*}
{\bf x}^{\prime }(g) &=&{\bf x}_{\parallel }^{\prime }(g)+{\bf x}_{\perp
}^{\prime }(g)=\left\{
\begin{array}{l}
{\bf x}+(\widetilde{\gamma }(g)-1)\frac{{\bf v}(g)}{\left| {\bf v}(g)\right|
_{\ast }^{2}}({\bf v}(g)\ast {\bf x})-\widetilde{\gamma }(g){\bf v}%
(g)t=\bigskip \\
\\
\\
={\bf x}+(\widetilde{\gamma }(g)-1)\frac{%
\sum_{k=1}^{3}b_{k}^{2}(x^{5})v^{k}(g)x^{k}}{\sum_{k=1}^{3}b_{k}^{2}(x^{5})%
\left( v^{k}(g)\right) ^{2}}{\bf v}(g)+ \\
\\
-\widetilde{\gamma }(g){\bf v}(g)t;\medskip
\end{array}
\right. \\
&&
\end{eqnarray*}
\begin{eqnarray}
t^{\prime }(g) &=&\left\{
\begin{array}{l}
\widetilde{\gamma }(g)(t-{\bf \tilde{B}}(g)\cdot {\bf x})=\widetilde{\gamma }%
(g)(t-{\bf \tilde{B}}^{(\ast )}(g)\ast {\bf x})= \\
\bigskip \\
=\widetilde{\gamma }(g)\left( t-\sum_{k=1}^{3}\frac{v^{k}(g)b_{k}^{2}(x^{5})%
}{c^{2}b_{0}^{2}(x^{5})}x^{k}\medskip \right) ,
\end{array}
\right.  \nonumber \\
&&
\end{eqnarray}
or (as usual $x^{0}\equiv ct$, and $\bullet $ now denotes usual algebraic
multiplication) ($\forall i=1,2,3$) :
\begin{gather*}
\left( x^{\prime }\right) _{(g),DSR4}^{0}(\left\{ x\right\}
_{m.},x^{5})=\left( x^{0}\right) _{(g),DSR4}^{^{\prime }}(\left\{ x\right\}
_{m.},x^{5})= \\
\\
=\widetilde{\gamma }(g)(x^{0}-\sum_{k=1}^{3}\frac{v^{k}(g)b_{k}^{2}(x^{5})}{%
cb_{0}^{2}(x^{5})}x^{k})= \\
\\
=\left( 1-\left( \frac{v^{1}(g)b_{1}(x^{5})}{cb_{0}(x^{5})}\right)
^{2}-\left( \frac{v^{2}(g)b_{2}(x^{5})}{cb_{0}(x^{5})}\right) ^{2}-\left(
\frac{v^{3}(g)b_{3}(x^{5})}{cb_{0}(x^{5})}\right) ^{2}\right) ^{-1/2}\bullet
\\
\\
\bullet \left( x^{0}-\sum_{k=1}^{3}\frac{v^{k}(g)b_{k}^{2}(x^{5})}{%
cb_{0}^{2}(x^{5})}x^{k}\right) \bigskip ;
\end{gather*}
\begin{gather}
\left( x^{\prime }\right) _{(g),DSR4}^{i}(\left\{ x\right\}
_{m.},x^{5})=\left( x^{i}\right) _{(g),DSR4}^{^{\prime }}(\left\{ x\right\}
_{m.},x^{5})=  \nonumber \\
\nonumber \\
=x^{i}+(\widetilde{\gamma }(g)-1)\frac{%
\sum_{k=1}^{3}b_{k}^{2}(x^{5})v^{k}(g)x^{k}}{\sum_{k=1}^{3}b_{k}^{2}(x^{5})%
\left( v^{k}(g)\right) ^{2}}v^{i}(g)-\widetilde{\gamma }(g)\frac{v^{i}(g)}{c}%
x^{0}=  \nonumber \\
\nonumber \\
=x^{i}+\bigskip  \nonumber \\
\nonumber \\
+\left( \left( 1-\left( \frac{v^{1}(g)b_{1}(x^{5})}{cb_{0}(x^{5})}\right)
^{2}-\left( \frac{v^{2}(g)b_{2}(x^{5})}{cb_{0}(x^{5})}\right) ^{2}-\left(
\frac{v^{3}(g)b_{3}(x^{5})}{cb_{0}(x^{5})}\right) ^{2}\right)
^{-1/2}-1\right) \bullet  \nonumber \\
\nonumber \\
\bullet \frac{\sum_{k=1}^{3}b_{k}^{2}(x^{5})v^{k}(g)x^{k}}{%
\sum_{k=1}^{3}b_{k}^{2}(x^{5})\left( v^{k}(g)\right) ^{2}}v^{i}(g)+\bigskip
\nonumber \\
\nonumber \\
-\frac{1}{c}\left( 1-\left( \frac{v^{1}(g)b_{1}(x^{5})}{cb_{0}(x^{5})}%
\right) ^{2}-\left( \frac{v^{2}(g)b_{2}(x^{5})}{cb_{0}(x^{5})}\right)
^{2}-\left( \frac{v^{3}(g)b_{3}(x^{5})}{cb_{0}(x^{5})}\right) ^{2}\right)
^{-1/2}x^{0}v^{i}(g)\bigskip .  \nonumber \\
\end{gather}
\qquad \qquad \qquad

Different explicit forms of a finite, deformed boost with generic velocity $%
{\bf v}(g)=v(g)\widehat{v}(g)\equiv v(g)\widehat{\varepsilon }(g)$ along a
generic direction $\widehat{\varepsilon }(g)$ in $\widetilde{E_{3}}%
(x^{5})\subset \widetilde{M_{4}}(x^{5})$, and consequently of $\exp \left(
B_{\left( \rho (g),\widehat{\varepsilon }(g)\right) ,DSR4}(x^{5})\right) $,
can be obtained by exploiting Eqs. (113)-(115).

Notice that the lack of symmetry properties of the $4\times 4$ matrices
representing deformed boosts is obviously related to the {\em %
''anisotropizing deforming''} character of the DSR4 generalization of SR.

\subsection{Parametric Change of Basis for a Deformed Boost in a Generic
Direction}

On account of Eqs. (55)-(57), which relate, through the use of definition
(44), the dimensionless parameter basis of deformed rapidities $\left\{
\zeta ^{i}(g)\right\} $, $\left\{ \tilde{\zeta}^{i}(g)\right\} $ and the
dimensional parameter basis of deformed boost velocities $\left\{
v^{i}(g)\right\} $, one gets (ESC off)
\begin{equation}
\left\{
\begin{array}{l}
I)\text{ \ }\frac{v^{i}(g)b_{i}(x^{5})}{cb_{0}(x^{5})}\equiv \widetilde{%
\beta }^{i}(g)=b_{i}(x^{5})\widetilde{\beta }^{i(\ast )}(g)= \\
\\
=tgh\left( \zeta _{i}(g)b_{0}^{-1}(x^{5})b_{i}^{-1}(x^{5})\right) =tgh\left(
\widetilde{\zeta }_{i}(g)\right) \bigskip ; \\
\\
\\
II)\text{ \ }\left( 1-\frac{b_{i}^{2}(x^{5})}{c^{2}b_{0}^{2}(x^{5})}\left(
v^{i}(g)\right) ^{2}\right) ^{-1/2}=(1-(\widetilde{\beta }%
^{i}(g))^{2})^{-1/2}= \\
\medskip \\
=(1-b_{i}^{2}(x^{5})(\widetilde{\beta }^{i(\ast )}(g))^{2})^{-1/2}\equiv
\widetilde{\gamma }^{i}(g)= \\
\\
=\cosh \left( \zeta _{i}(g)b_{0}^{-1}(x^{5})b_{i}^{-1}(x^{5})\right)
\medskip =\cosh \left( \widetilde{\zeta }_{i}(g)\right) ,
\end{array}
\right.
\end{equation}
and therefore
\begin{gather}
\widetilde{\gamma }(g)\equiv  \nonumber \\
\nonumber \\
\nonumber \\
\equiv \left( 1-\frac{1}{c^{2}b_{0}^{2}(x^{5})}\sum_{k=1}^{3}\left(
v^{i}(g)\right) ^{2}b_{i}^{2}(x^{5})\right) ^{-1/2}=\left( 1-\sum_{i=1}^{3}(%
\widetilde{\beta }^{i}(g))^{2}\right) ^{-1/2}=  \nonumber \\
\nonumber \\
\nonumber \\
=\left( 1-\sum_{i=1}^{3}b_{i}^{2}(x^{5})(\widetilde{\beta }^{i(\ast
)}(g))^{2}\right) ^{-1/2}=\left( 1-\sum_{i=1}^{3}(tgh\left( \zeta
_{i}(g)b_{0}^{-1}(x^{5})b_{i}^{-1}(x^{5})\right) )^{2}\right) ^{-1/2}=
\nonumber \\
\nonumber \\
\nonumber \\
=\left( 1-\sum_{i=1}^{3}(tgh\left( \widetilde{\zeta }_{i}(g)\right)
)^{2}\right) ^{-1/2};  \nonumber \\
\\
\nonumber
\end{gather}
\begin{gather}
\frac{\left( \widetilde{\beta ^{i}}(g)\right) ^{2}}{|{\bf \tilde{\beta}}%
(g)|^{2}}=\frac{b_{i}^{2}(x^{5})\left( \widetilde{\beta ^{i}}^{(\ast
)}(g)\right) ^{2}}{|{\bf \tilde{\beta}}^{(\ast )}(g)|_{\ast }^{2}}=
\nonumber \\
\nonumber \\
\nonumber \\
=\frac{\left( \widetilde{\beta ^{i}}(g)\right) ^{2}}{{\bf \tilde{\beta}}%
(g)\cdot {\bf \tilde{\beta}}(g)}=\frac{b_{i}^{2}(x^{5})\left( \widetilde{%
\beta ^{i}}^{(\ast )}(g)\right) ^{2}}{{\bf \tilde{\beta}}^{(\ast )}(g)\ast
{\bf \tilde{\beta}}^{(\ast )}(g)}=  \nonumber \\
\nonumber \\
\nonumber \\
=\frac{\left( \widetilde{\beta ^{i}}(g)\right) ^{2}}{\sum_{k=1}^{3}(%
\widetilde{\beta ^{k}}(g))^{2}}=\frac{b_{i}^{2}(x^{5})\left( \widetilde{%
\beta ^{i}}^{(\ast )}(g)\right) ^{2}}{\sum_{k=1}^{3}b_{k}^{2}(x^{5})(%
\widetilde{\beta ^{k}}^{(\ast )}(g))^{2}}=  \nonumber \\
\nonumber \\
\nonumber \\
=\frac{\left( tgh\left( \zeta
_{i}(g)b_{0}^{-1}(x^{5})b_{i}^{-1}(x^{5})\right) \right) ^{2}}{%
\sum_{k=1}^{3}(tgh\left( \zeta
_{k}(g)b_{0}^{-1}(x^{5})b_{k}^{-1}(x^{5})\right) )^{2}}=  \nonumber \\
\nonumber \\
\nonumber \\
=\frac{\left( tgh\left( \widetilde{\zeta }_{i}(g)\right) \right) ^{2}}{%
\sum_{k=1}^{3}(tgh\left( \widetilde{\zeta }_{k}(g)\right) )^{2}};  \nonumber
\\
\\
\nonumber
\end{gather}
\begin{gather}
\frac{\widetilde{\beta ^{i}}(g)\widetilde{\beta ^{j}}(g)}{|{\bf \tilde{\beta}%
}(g)|^{2}}=\frac{b_{i}(x^{5})b_{j}(x^{5})\widetilde{\beta ^{i}}^{(\ast )}(g)%
\widetilde{\beta ^{j}}^{(\ast )}(g)}{|{\bf \tilde{\beta}}^{(\ast
)}(g)|_{\ast }^{2}}=  \nonumber \\
\nonumber \\
\nonumber \\
=\frac{\widetilde{\beta ^{i}}(g)\widetilde{\beta ^{j}}(g)}{{\bf \tilde{\beta}%
}(g)\cdot {\bf \tilde{\beta}}(g)}=\frac{b_{i}(x^{5})b_{j}(x^{5})\widetilde{%
\beta ^{i}}^{(\ast )}(g)\widetilde{\beta ^{j}}^{(\ast )}(g)}{{\bf \tilde{%
\beta}}^{(\ast )}(g)\ast {\bf \tilde{\beta}}^{(\ast )}(g)}=  \nonumber \\
\nonumber \\
\nonumber \\
=\frac{\widetilde{\beta ^{i}}(g)\widetilde{\beta ^{j}}(g)}{\sum_{k=1}^{3}(%
\widetilde{\beta }^{k}(g))^{2}}=\frac{b_{i}(x^{5})b_{j}(x^{5})\widetilde{%
\beta ^{i}}^{(\ast )}(g)\widetilde{\beta ^{j}}^{(\ast )}(g)}{%
\sum_{k=1}^{3}b_{k}^{2}(x^{5})(\widetilde{\beta }^{k(\ast )}(g))^{2}}=
\nonumber \\
\nonumber \\
\nonumber \\
=\frac{\left( tgh\left( \zeta
_{i}(g)b_{0}^{-1}(x^{5})b_{i}^{-1}(x^{5})\right) \right) \left( tgh\left(
\zeta _{j}(g)b_{0}^{-1}(x^{5})b_{j}^{-1}(x^{5})\right) \right) }{%
\sum_{k=1}^{3}(tgh\left( \zeta
_{k}(g)b_{0}^{-1}(x^{5})b_{k}^{-1}(x^{5})\right) )^{2}}=  \nonumber \\
\nonumber \\
\nonumber \\
=\frac{\left( tgh\left( \widetilde{\zeta }_{i}(g)\right) \right) \left(
tgh\left( \widetilde{\zeta }_{j}(g)\right) \right) }{\sum_{k=1}^{3}(tgh%
\left( \widetilde{\zeta }_{k}(g)\right) )^{2}};  \nonumber \\
\end{gather}
\begin{gather}
\widetilde{\beta }^{2}\equiv |{\bf \tilde{\beta}}|^{2}=|{\bf \tilde{\beta}}%
^{(\ast )}|_{\ast }^{2}={\bf \tilde{\beta}}(g)\cdot {\bf \tilde{\beta}}(g)=
\nonumber \\
\nonumber \\
\nonumber \\
={\bf \tilde{\beta}}^{(\ast )}(g)\ast {\bf \tilde{\beta}}^{(\ast )}(g)=
\nonumber \\
\nonumber \\
\nonumber \\
=\sum_{k=1}^{3}(\widetilde{\beta }^{k}(g))^{2}=%
\sum_{k=1}^{3}b_{k}^{2}(x^{5})(\widetilde{\beta }^{k(\ast )}(g))^{2}=
\nonumber \\
\nonumber \\
\nonumber \\
=\sum_{k=1}^{3}(tgh\left( \zeta
_{k}(g)b_{0}^{-1}(x^{5})b_{k}^{-1}(x^{5})\right) )^{2}=  \nonumber \\
\nonumber \\
\nonumber \\
=\sum_{k=1}^{3}(tgh\left( \widetilde{\zeta }_{k}(g)\right) )^{2}.  \nonumber
\\
\\
\nonumber
\end{gather}

The above relations between the sets $\left\{ \zeta ^{i}(g)\right\}
_{i=1,2,3}$ ,$\left\{ \tilde{\zeta}^{i}(g)\right\} _{i=1,2,3}$and $\left\{
v^{i}(g)\right\} _{i=1,2,3}$ express the change of parametric base for
deformed boosts in DSR4, from the dimensional parameter basis of deformed
boost velocities to the dimensionless parameter basis of deformed
rapidities, By means of Eqs. (122)-(125) one can therefore write $\exp
\left( B_{\left( \rho (g),\widehat{\varepsilon }(g)\right)
,DSR4}(x^{5})\right) $ (derived in the previous Subsection) in terms of
rapidities. We have for instance, in terms of the effective deformed
rapidities{\bf \ }$\widetilde{\zeta }^{i}(g,x^{5})$ defined by Eq. (44),
\begin{eqnarray}
1)\text{ \ }\left( \exp \left( B_{\left( \rho (g),(\widehat{v}(g)\equiv
\widehat{\varepsilon }(g)\right) ,DSR4}(x^{5})\right) \right) _{~~0}^{0}
&=&\left( 1-\sum_{i=1}^{3}(tgh\left( \tilde{\zeta}^{i}(g,x^{5})\right)
)^{2}\right) ^{-1/2}\medskip \medskip ;  \nonumber \\
&&
\end{eqnarray}
\begin{gather}
2)\text{ \ }\left( \exp \left( B_{\left( \rho (g),(\widehat{v}(g)\equiv
\widehat{\varepsilon }(g)\right) ,DSR4}(x^{5})\right) \right)
_{~~m}^{0}=\medskip  \nonumber \\
\nonumber \\
\nonumber \\
=-\left( 1-\sum_{i=1}^{3}(tgh\left( \tilde{\zeta}^{i}(g,x^{5})\right)
)^{2}\right) ^{-1/2}\frac{b_{m}(x^{5})}{b_{0}(x^{5})}tgh\left( \tilde{\zeta}%
^{m}(g,x^{5})\right) ,\text{ }\forall m=1,2,3\medskip \medskip ;  \nonumber
\\
\end{gather}
\begin{gather}
3)\text{ \ }\left( \exp \left( B_{\left( \rho (g),(\widehat{v}(g)\equiv
\widehat{\varepsilon }(g)\right) ,DSR4}(x^{5})\right) \right)
_{~~0}^{m}=\medskip  \nonumber \\
\nonumber \\
\nonumber \\
=-\left( 1-\sum_{i=1}^{3}(tgh\left( \tilde{\zeta}^{i}(g,x^{5})\right)
)^{2}\right) ^{-1/2}\frac{b_{0}(x^{5})}{b_{m}(x^{5})}tgh\left( \tilde{\zeta}%
^{m}(g,x^{5})\right) ,\text{ }\forall m=1,2,3\medskip ;  \nonumber \\
\end{gather}
\begin{gather}
4)\text{ \ }\left( \exp \left( B_{\left( \rho (g),(\widehat{v}(g)\equiv
\widehat{\varepsilon }(g)\right) ,DSR4}(x^{5})\right) \right)
_{~~m}^{m}=\medskip  \nonumber \\
\nonumber \\
\nonumber \\
=1+\left( \left( 1-\sum_{i=1}^{3}(tgh\left( \tilde{\zeta}^{i}(g,x^{5})%
\right) )^{2}\right) ^{-1/2}-1\right) \frac{\left( tgh\left( \tilde{\zeta}%
^{m}(g,x^{5})\right) \right) ^{2}}{\sum_{k=1}^{3}(tgh\left( \tilde{\zeta}%
^{k}(g,x^{5})\right) )^{2}},\text{ }\forall m=1,2,3  \nonumber \\
\end{gather}
\begin{gather}
5)\text{ \ }\left( \exp \left( B_{\left( \rho (g),(\widehat{v}(g)\equiv
\widehat{\varepsilon }(g)\right) ,DSR4}(x^{5})\right) \right)
_{~~2}^{1}=\medskip  \nonumber \\
\nonumber \\
\nonumber \\
=\left( \left( 1-\sum_{i=1}^{3}(tgh\left( \tilde{\zeta}^{i}(g,x^{5})\right)
)^{2}\right) ^{-1/2}-1\right) \frac{b_{2}(x^{5})}{b_{1}(x^{5})}\frac{\left(
tgh\left( \tilde{\zeta}^{1}(g,x^{5})\right) \right) \left( tgh\left( \tilde{%
\zeta}^{2}(g,x^{5})\right) \right) }{\sum_{k=1}^{3}(tgh\left( \tilde{\zeta}%
^{k}(g,x^{5})\right) )^{2}};  \nonumber \\
\end{gather}
\begin{gather}
6)\text{ \ }\left( \exp \left( B_{\left( \rho (g),(\widehat{v}(g)\equiv
\widehat{\varepsilon }(g)\right) ,DSR4}(x^{5})\right) \right)
_{~~3}^{1}=\medskip  \nonumber \\
\nonumber \\
\nonumber \\
=\left( \left( 1-\sum_{i=1}^{3}(tgh\left( \tilde{\zeta}^{i}(g,x^{5})\right)
)^{2}\right) ^{-1/2}-1\right) \frac{b_{3}(x^{5})}{b_{1}(x^{5})}\frac{\left(
tgh\left( \tilde{\zeta}^{1}(g,x^{5})\right) \right) \left( tgh\left( \tilde{%
\zeta}^{3}(g,x^{5})\right) \right) }{\sum_{k=1}^{3}(tgh\left( \tilde{\zeta}%
^{k}(g,x^{5})\right) )^{2}};  \nonumber \\
\end{gather}
\begin{gather}
7)\text{ \ }\left( \exp \left( B_{\left( \rho (g),(\widehat{v}(g)\equiv
\widehat{\varepsilon }(g)\right) ,DSR4}(x^{5})\right) \right)
_{~~1}^{2}=\medskip  \nonumber \\
\nonumber \\
\nonumber \\
=\left( \left( 1-\sum_{i=1}^{3}(tgh\left( \tilde{\zeta}^{i}(g,x^{5})\right)
)^{2}\right) ^{-1/2}-1\right) \frac{b_{1}(x^{5})}{b_{2}(x^{5})}\frac{\left(
tgh\left( \tilde{\zeta}^{1}(g,x^{5})\right) \right) \left( tgh\left( \tilde{%
\zeta}^{2}(g,x^{5})\right) \right) }{\sum_{k=1}^{3}(tgh\left( \tilde{\zeta}%
^{k}(g,x^{5})\right) )^{2}};  \nonumber \\
\end{gather}
\begin{gather}
8)\text{ \ }\left( \exp \left( B_{\left( \rho (g),(\widehat{v}(g)\equiv
\widehat{\varepsilon }(g)\right) ,DSR4}(x^{5})\right) \right)
_{~~3}^{2}=\medskip  \nonumber \\
\nonumber \\
\nonumber \\
=\left( \left( 1-\sum_{i=1}^{3}(tgh\left( \tilde{\zeta}^{i}(g,x^{5})\right)
)^{2}\right) ^{-1/2}-1\right) \frac{b_{3}(x^{5})}{b_{2}(x^{5})}\frac{\left(
tgh\left( \tilde{\zeta}^{2}(g,x^{5})\right) \right) \left( tgh\left( \tilde{%
\zeta}^{3}(g,x^{5})\right) \right) }{\sum_{k=1}^{3}(tgh\left( \tilde{\zeta}%
^{k}(g,x^{5})\right) )^{2}};  \nonumber \\
\end{gather}
\begin{gather}
9)\text{ \ }\left( \exp \left( B_{\left( \rho (g),(\widehat{v}(g)\equiv
\widehat{\varepsilon }(g)\right) ,DSR4}(x^{5})\right) \right)
_{~~1}^{3}=\medskip  \nonumber \\
\nonumber \\
\nonumber \\
=\left( \left( 1-\sum_{i=1}^{3}(tgh\left( \tilde{\zeta}^{i}(g,x^{5})\right)
)^{2}\right) ^{-1/2}-1\right) \frac{b_{1}(x^{5})}{b_{3}(x^{5})}\frac{\left(
tgh\left( \tilde{\zeta}^{1}(g,x^{5})\right) \right) \left( tgh\left( \tilde{%
\zeta}^{3}(g,x^{5})\right) \right) }{\sum_{k=1}^{3}(tgh\left( \tilde{\zeta}%
^{k}(g,x^{5})\right) )^{2}};  \nonumber \\
\end{gather}
\begin{gather}
10)\text{ \ }\left( \exp \left( B_{\left( \rho (g),(\widehat{v}(g)\equiv
\widehat{\varepsilon }(g)\right) ,DSR4}(x^{5})\right) \right)
_{~~2}^{3}=\medskip  \nonumber \\
\nonumber \\
\nonumber \\
=\left( \left( 1-\sum_{i=1}^{3}(tgh\left( \tilde{\zeta}^{i}(g,x^{5})\right)
)^{2}\right) ^{-1/2}-1\right) \frac{b_{2}(x^{5})}{b_{3}(x^{5})}\frac{\left(
tgh\left( \tilde{\zeta}^{2}(g,x^{5})\right) \right) \left( tgh\left( \tilde{%
\zeta}^{3}(g,x^{5})\right) \right) }{\sum_{k=1}^{3}(tgh\left( \tilde{\zeta}%
^{k}(g,x^{5})\right) )^{2}};  \nonumber \\
\end{gather}
\begin{gather}
\left( x^{\prime }\right) _{(g),DSR4}^{0}(\left\{ x\right\}
_{m.},x^{5})=\left( x^{0}\right) _{(g),DSR4}^{^{\prime }}(\left\{ x\right\}
_{m.},x^{5})=  \nonumber \\
\nonumber \\
\nonumber \\
=\medskip \sum_{\mu =0}^{3}\left( \exp \left( B_{\left( \rho (g),\widehat{v}%
(g)\equiv \widehat{\varepsilon }(g)\right) ,DSR4}(x^{5})\right) \right)
_{~~\mu }^{0}x^{\mu }=  \nonumber \\
\nonumber \\
\nonumber \\
=\left[ \left( 1-\sum_{i=1}^{3}(tgh\left( \tilde{\zeta}_{i}(g,x^{5})\right)
)^{2}\right) ^{-1/2}\right] x^{0}+\medskip  \nonumber \\
\nonumber \\
\nonumber \\
-\left[ \left( 1-\sum_{i=1}^{3}(tgh\left( \tilde{\zeta}_{i}(g,x^{5})\right)
)^{2}\right) ^{-1/2}\frac{b_{1}(x^{5})}{b_{0}(x^{5})}tgh\left( \tilde{\zeta}%
_{1}(g,x^{5})\right) \right] x^{1}+\medskip  \nonumber \\
\nonumber \\
\nonumber \\
-\medskip \bigskip \left[ \left( 1-\sum_{i=1}^{3}(tgh\left( \tilde{\zeta}%
_{i}(g,x^{5})\right) )^{2}\right) ^{-1/2}\frac{b_{2}(x^{5})}{b_{0}(x^{5})}%
tgh\left( \tilde{\zeta}_{2}(g,x^{5})\right) \right] x^{2}+  \nonumber \\
\medskip  \nonumber \\
\nonumber \\
-\left[ \left( 1-\sum_{i=1}^{3}(tgh\left( \tilde{\zeta}_{i}(g,x^{5})\right)
)^{2}\right) ^{-1/2}\frac{b_{3}(x^{5})}{b_{0}(x^{5})}tgh\left( \tilde{\zeta}%
_{3}(g,x^{5})\right) \right] x^{3}\medskip ;  \nonumber \\
\end{gather}
\begin{gather}
\left( x^{\prime }\right) _{(g),DSR4}^{1}(\left\{ x\right\}
_{m.},x^{5})=\left( x^{1}\right) _{(g),DSR4}^{^{\prime }}(\left\{ x\right\}
_{m.},x^{5})=  \nonumber \\
\nonumber \\
\nonumber \\
=\medskip \bigskip \sum_{\mu =0}^{3}\left( \exp \left( B_{\left( \rho (g),%
\widehat{v}(g)\equiv \widehat{\varepsilon }(g)\right) ,DSR4}(x^{5})\right)
\right) _{~~\mu }^{1}x^{\mu }=  \nonumber \\
\nonumber \\
\nonumber \\
=-\left[ \left( 1-\sum_{i=1}^{3}(tgh\left( \tilde{\zeta}_{i}(g,x^{5})\right)
)^{2}\right) ^{-1/2}\frac{b_{0}(x^{5})}{b_{1}(x^{5})}tgh\left( \tilde{\zeta}%
_{1}(g,x^{5})\right) \right] x^{0}+\medskip  \nonumber \\
\nonumber \\
\nonumber \\
+\left[ 1+\left( \left( 1-\sum_{i=1}^{3}(\tilde{\zeta}_{i}(g,x^{5}))^{2}%
\right) ^{-1/2}-1\right) \frac{\left( tgh\left( \tilde{\zeta}%
_{1}(g,x^{5})\right) \right) ^{2}}{\sum_{k=1}^{3}(tgh\left( \tilde{\zeta}%
_{k}(g,x^{5})\right) )^{2}}\right] x^{1}+  \nonumber \\
\medskip  \nonumber \\
\nonumber \\
+\left[ \left( \left( 1-\sum_{i=1}^{3}(tgh\left( \tilde{\zeta}%
_{i}(g,x^{5})\right) )^{2}\right) ^{-1/2}-1\right) \bullet \right.  \nonumber
\\
\nonumber \\
\left. \bullet \frac{b_{2}(x^{5})}{b_{1}(x^{5})}\frac{\left( tgh\left(
\tilde{\zeta}_{1}(g,x^{5})\right) \right) \left( tgh\left( \tilde{\zeta}%
_{2}(g,x^{5})\right) \right) }{\sum_{k=1}^{3}(tgh\left( \tilde{\zeta}%
_{k}(g,x^{5})\right) )^{2}}\right] x^{2}+\medskip  \nonumber \\
\nonumber \\
\nonumber \\
+\left[ \left( \left( 1-\sum_{i=1}^{3}(tgh\left( \tilde{\zeta}%
_{i}(g,x^{5})\right) )^{2}\right) ^{-1/2}-1\right) \bullet \right.  \nonumber
\\
\nonumber \\
\left. \bullet \frac{b_{3}(x^{5})}{b_{1}(x^{5})}\frac{\left( tgh\left(
\tilde{\zeta}_{1}(g,x^{5})\right) \right) \left( tgh\left( \tilde{\zeta}%
_{3}(g,x^{5})\right) \right) }{\sum_{k=1}^{3}(tgh\left( \tilde{\zeta}%
_{k}(g,x^{5})\right) )^{2}}\right] x^{3}\medskip ;  \nonumber \\
\end{gather}
\begin{gather}
\left( x^{\prime }\right) _{(g),DSR4}^{2}(\left\{ x\right\}
_{m.},x^{5})=\left( x^{2}\right) _{(g),DSR4}^{^{\prime }}(\left\{ x\right\}
_{m.},x^{5})=  \nonumber \\
\nonumber \\
\nonumber \\
=\medskip \bigskip \sum_{\mu =0}^{3}\left( \exp \left( B_{\left( \rho (g),%
\widehat{v}(g)\equiv \widehat{\varepsilon }(g)\right) ,DSR4}(x^{5})\right)
\right) _{~~\mu }^{2}x^{\mu }=  \nonumber \\
\nonumber \\
\nonumber \\
=-\left[ \left( 1-\sum_{i=1}^{3}(tgh\left( \tilde{\zeta}_{i}(g,x^{5})\right)
)^{2}\right) ^{-1/2}\frac{b_{0}(x^{5})}{b_{2}(x^{5})}tgh\left( \tilde{\zeta}%
_{2}(g,x^{5})\right) \right] x^{0}+\medskip  \nonumber \\
\nonumber \\
\nonumber \\
+\left[ \left( \left( 1-\sum_{i=1}^{3}(tgh\left( \tilde{\zeta}%
_{i}(g,x^{5})\right) )^{2}\right) ^{-1/2}-1\right) \bullet \right.  \nonumber
\\
\nonumber \\
\left. \bullet \frac{b_{1}(x^{5})}{b_{2}(x^{5})}\frac{\left( tgh\left(
\tilde{\zeta}_{1}(g,x^{5})\right) \right) \left( tgh\left( \tilde{\zeta}%
_{2}(g,x^{5})\right) \right) }{\sum_{k=1}^{3}(tgh\left( \tilde{\zeta}%
_{k}(g,x^{5})\right) )^{2}}\right] x^{1}+\medskip  \nonumber \\
\nonumber \\
+\left[ 1+\left( \left( 1-\sum_{i=1}^{3}(tgh\left( \tilde{\zeta}%
_{i}(g,x^{5})\right) )^{2}\right) ^{-1/2}-1\right) \frac{\left( tgh\left(
\tilde{\zeta}_{2}(g,x^{5})\right) \right) ^{2}}{\sum_{k=1}^{3}(tgh\left(
\tilde{\zeta}_{k}(g,x^{5})\right) )^{2}}\right] x^{2}+  \nonumber \\
\medskip  \nonumber \\
\nonumber \\
+\left[ \left( \left( 1-\sum_{i=1}^{3}(tgh\left( \tilde{\zeta}%
_{i}(g,x^{5})\right) )^{2}\right) ^{-1/2}-1\right) \bullet \right.  \nonumber
\\
\nonumber \\
\left. \bullet \frac{b_{3}(x^{5})}{b_{2}(x^{5})}\frac{\left( tgh\left(
\tilde{\zeta}_{2}(g,x^{5})\right) \right) \left( tgh\left( \tilde{\zeta}%
_{3}(g,x^{5})\right) \right) }{\sum_{k=1}^{3}(tgh\left( \tilde{\zeta}%
_{k}(g,x^{5})\right) )^{2}}\right] x^{3}\medskip ;  \nonumber \\
\end{gather}
\begin{gather}
\left( x^{\prime }\right) _{(g),DSR4}^{3}(\left\{ x\right\}
_{m.},x^{5})=\left( x^{3}\right) _{(g),DSR4}^{^{\prime }}(\left\{ x\right\}
_{m.},x^{5})=  \nonumber \\
\nonumber \\
\nonumber \\
=\medskip \bigskip \sum_{\mu =0}^{3}\left( \exp \left( B_{\left( \rho (g),%
\widehat{v}(g)\equiv \widehat{\varepsilon }(g)\right) ,DSR4}(x^{5})\right)
\right) _{~~\mu }^{3}x^{\mu }=  \nonumber \\
\nonumber \\
\nonumber \\
=-\left[ \left( 1-\sum_{i=1}^{3}(tgh\left( \tilde{\zeta}_{i}(g,x^{5})\right)
)^{2}\right) ^{-1/2}\frac{b_{0}(x^{5})}{b_{3}(x^{5})}tgh\left( \tilde{\zeta}%
_{3}(g,x^{5})\right) \right] x^{0}+\medskip  \nonumber \\
\nonumber \\
\nonumber \\
+\left[ \left( \left( 1-\sum_{i=1}^{3}(tgh\left( \tilde{\zeta}%
_{i}(g,x^{5})\right) )^{2}\right) ^{-1/2}-1\right) \bullet \right.  \nonumber
\\
\nonumber \\
\left. \bullet \frac{b_{1}(x^{5})}{b_{3}(x^{5})}\frac{\left( tgh\left(
\tilde{\zeta}_{1}(g,x^{5})\right) \right) \left( tgh\left( \tilde{\zeta}%
_{3}(g,x^{5})\right) \right) }{\sum_{k=1}^{3}(tgh\left( \tilde{\zeta}%
_{k}(g,x^{5})\right) )^{2}}\right] x^{1}+\medskip  \nonumber \\
\nonumber \\
\nonumber \\
+\left[ \left( \left( 1-\sum_{i=1}^{3}(tgh(\tilde{\zeta}_{i}(g,x^{5})))^{2}%
\right) ^{-1/2}-1\right) \bullet \right.  \nonumber \\
\nonumber \\
\left. \bullet \frac{b_{2}(x^{5})}{b_{3}(x^{5})}\frac{\left( tgh\left(
\tilde{\zeta}_{2}(g,x^{5})\right) \right) \left( tgh\left( \tilde{\zeta}%
_{3}(g,x^{5})\right) \right) }{\sum_{k=1}^{3}(tgh\left( \tilde{\zeta}%
_{k}(g,x^{5})\right) )^{2}}\right] x^{2}+\medskip  \nonumber \\
\nonumber \\
\nonumber \\
+\left[ 1+\left( \left( 1-\sum_{i=1}^{3}(tgh\left( \tilde{\zeta}%
_{i}(g,x^{5})\right) )^{2}\right) ^{-1/2}-1\right) \frac{\left( tgh\left(
\tilde{\zeta}_{3}(g,x^{5})\right) \right) ^{2}}{\sum_{k=1}^{3}(tgh\left(
\tilde{\zeta}_{k}(g,x^{5})\right) )^{2}}\right] x^{3}\medskip .  \nonumber \\
\end{gather}

\section{CONCLUSIONS}

In this paper, we discussed the finite structure of the space-time\
rotations in generalized Minkowski spaces $\widetilde{M_{N}}(\left\{
x\right\} _{n.m.})$, i.e. $N$-dimensional spaces endowed with a (in general
non-diagonal) metric tensor, whose coefficients do depend on a set of
non-metrical coordinates. We considered in detail (without loss of
generality) the four-dimensional case. In particular, the results obtained
have been specialized to the case of a ''deformed'' Minkowski space $%
\widetilde{M_{4}}$ (i.e. a pseudoeuclidean space with metric coefficients
depending on energy), for which we derived the explicit general form of the
finite rotations and boosts in different parametric bases.

This concludes the study (started in Ref. [1]) of the rotational component
of the maximal Killing group of $\widetilde{M_{N}}(\left\{ x\right\}
_{n.m.}) $, namely of the $N$-d. generalized, homogeneous Lorentz group $%
SO(T,S)_{GEN.}^{N(N-1)/2\text{{\footnotesize \ }}}$, subgroup of the
generalized Poincar\'{e} group $P(S,T)_{GEN.}^{N(N+1)/2}$. The space-time
translations in generalized Minkowski spaces will be investigated in a
forthcoming paper.\pagebreak

\ \

\bigskip

\end{document}